\documentclass{svjour2}                  
\usepackage{graphicx,amsmath}
 \usepackage{mathptmx,color}     
\usepackage{latexsym}
\begin{document}

\title{Two refreshing views of Fluctuation Theorems through Kinematics Elements and Exponential
Martingale}
\author{Rapha\"{e}l Chetrite \and
        Shamik Gupta.
}

\institute{Rapha\"{e}l Chetrite \and Shamik Gupta \at
              Physics of Complex Systems, Weizmann Institute of Science, Rehovot 76100, Israel \\
           \and
           Rapha\"{e}l Chetrite \at
           Present address: Laboratoire J. A. Dieudonn\'{e}, UMR CNRS 6621,
Universit\'{e}e de Nice Sophia-Antipolis,
Parc Valrose, F-06108 Nice Cedex 02, France. \\ 
                            \email{raphael.chetrite@unice.fr}             \\
           \and
           Shamik Gupta \at
           Present address: Laboratoire de Physique de l'\'{E}cole Normale Sup\'{e}rieure de Lyon, Universit\'{e} de Lyon, CNRS, 46 All\'{e}e d'Italie, 69364 Lyon c\'{e}dex 07, France. \\ 
                            \email{shamik.gupta@ens-lyon.fr} 
}

\date{Received: date / Accepted: date}

\maketitle

\begin{abstract}
In the context of Markov  
evolution, we present two original approaches 
to obtain 
Generalized Fluctuation-Dissipation Theorems (\textbf{GFDT}), by
using the language of 
stochastic derivatives and by using a family of 
exponential martingales functionals. We show that \textbf{GFDT} are 
perturbative versions of relations verified by these 
exponential martingales. Along the way, we prove \textbf{GFDT} and 
Fluctuation Relations (\textbf{FR}) for general 
Markov processes, beyond the usual proof for diffusion 
and pure jump processes. Finally, we relate the \textbf{FR} to a family 
of backward and forward exponential martingales. 
\end{abstract}

\section{Introduction}

One of the cornerstones of statistical physics is the 
Fluctuation-Dissipation Theorem (\textbf{FDT}) \cite{Call1,Kubo0,MPRV,Risk}, 
whereby, for 
equilibrium systems, response to a small perturbation of the Hamiltonian is 
related to dynamical correlation. This theorem rationalizes the famous 
regression principle of Onsager \cite{Ons1,Ons2}: the decay of spontaneous 
fluctuation cannot be distinguished from the decay of forced fluctuation. 
More precisely, suppose we perturb a system in 
equilibrium at temperature $T$ by adding to its time-independent Hamiltonian $H$ a small time-dependent term, such that $H\rightarrow H-k_{t}O$. Here, $O$ is an observable and $k_{t}$ is a real function. Throughout this paper, we measure temperatures in units of the Boltzmann constant $k_B$. The \textbf{FDT} asserts that the response of an observable $O'$ is related to the two-time correlation function as 
\begin{equation}
T\left. \frac{\delta \left\langle O'_{t}\right\rangle ^{\prime }}{\delta k_{s}
}\right| _{k=0}=\partial _{s}\left\langle O_{s}O'_{t}\right\rangle ,
\label{ufdt}
\end{equation}
with $t\geq s$. In this relation, the brackets, 
$\left\langle{\ }\right\rangle 
$ and $\left\langle {\ }\right\rangle^{\prime }$, denote expectation in the
unperturbed and perturbed processes, respectively. Since mid-nineties,
this theorem has been extended to 
nonequilibrium systems in two
related directions. The first is the discovery of various 
Fluctuation Relations (\textbf{FR}) \cite{Crooks2,Eva,Gal,Jarz}, the 
so-called \textbf{
Gallavotti-Cohen relation} \cite{Eva,Gal}, the \textbf{
Jarzynski equality} \cite
{Jarz} and the \textbf{
Crooks theorem} \cite{Crooks2}. All of these hold
arbitrarily far from 
equilibrium and can be viewed as
non-perturbative extensions \cite{Gall2} of the \textbf{FDT} (\ref
{ufdt}). These relations constrain the distribution of entropy production or
work performed in the system. The second is the extension of the
relation (\ref{ufdt}) between response and correlation in the linear response 
regime to 
nonequilibrium states (stationary as well as non-stationary),
for example, those in glassy systems and soft spin models \cite
{Cri1,Cug1,Die1,Lip1,May1} and also in relation to broken supersymmetry \cite{Zim1}
. This second direction has seen an upsurge in the last three years through
formulation of the 
Generalized Fluctuation-Dissipation Theorems (\textbf{
GFDT}), mainly in the works of Seifert and Speck in Stuttgart \cite
{Sei2,SpS0,Spe2}, Baiesi, Maes and Wynants in Leuven \cite{Bae1,Bae2,Mae3},
and Gawedzki and Chetrite in Lyon and Falkovich in Rehovot \cite{Fal2,Che3,Che4} (see, also, the works 
\cite{Liu2,Pro1}). Moreover, experimental verifications of the {\bf GFDT} on colloidal particle
have been done in Lyon \cite{Gom1,Gom2}.

In the present paper, 
we revisit, generalize, and unify these \textbf{FR} and
\textbf{GFDT} by couching them in the language of the kinematics of a general Markov
process, without strict mathematical rigor. We show that this language allows
elementary proofs and generalizations of the different {\textbf GFDT} which 
exist in
the literature. We also consider a new family of non-perturbative extensions
of the {\textbf GFDT} which concerns a so-called \textbf{
exponential forward martingale} functionals 
\cite{Chu1,Rev1,Pal1}. 
Finally, we revisit the \textbf{FR} and show their relation to 
forward and backward exponential martingales\footnote{In the following, unless stated otherwise, use of the word martingale alone would mean forward martingale.}. In the process, we prove
that a certain version of the 
Crooks theorem and 
the Jarzynski equality hold for fairly
general Markov  
processes, whereas
the 
Gallavotti-Cohen relation for the performed work can be violated 
when the particle is subjected to a Poisson or Levy noise \cite{Bau1,Tou1}. 

General Markov  
stochastic processes form an integral part of modeling of
dynamics in statistical mechanics. Although largely idealized, they often provide a
sufficiently realistic description of experimental situations and have
traditionally served as a playground for both theoretical considerations and
numerical studies. In a continuous space (e.g., $\mathcal{R}^{d}$), all 
continuous time Markov processes consist of some combinations of diffusion,
deterministic motion and random jumps. Markov processes corresponding to
equilibrium 
dynamics are characterized by the detailed balance
property which ensures that the net probability flux between microstates of
the system vanishes. On the other hand, with 
nonequilibrium Markov
dynamics, detailed balance is violated and there are non-zero probability
fluxes even in a stationary situation. For the purpose of characterizing the
difference between 
equilibrium and 
nonequilibrium
dynamics, it is interesting to find a vector field, a kind of velocity, 
which vanishes in 
equilibrium. Such an object was introduced in
the sixties by Nelson in his seminal work \cite{Nel} with the notion of
current velocity that we call here the \textbf{local symmetric velocity}. This
quantity is an average of a well-chosen instantaneous velocity of the process
conditioned to pass through a given point. It was shown in 
\cite{Che4} that 
nonequilibrium diffusive dynamics (without the random jumps) takes, in
the Lagrangian frame of this velocity, an 
equilibrium form with the
detailed balance property and this explains the usual form (\ref{ufdt}) of the 
\textbf{FDT
} in that frame, which was observed previously in \cite{Fal2}. 
The issue regarding the extension of
this result to other types of Markov processes is addressed in this article 
in one of the Sections.

The formulation of the usual \textbf{FDT} (\ref{ufdt}) for some Markov
processes is known since long time \cite{Call1,Kubo0,Risk} in physics, but now it
has a strictly mathematically rigorous formulation \cite{Dem1}. For the 
\textbf{FR}, shortly after the earliest articles in the context of deterministic
dynamics \cite{Eva,Gal}, the fluctuation relations were proved for some
Markovian dynamics. In \cite{Jarz2}, Jarzynski generalized his relation to
time-dependent pure jump Markov processes. At around the same time, Kurchan showed in \cite{Kurchan} that the stationary \textbf{FR} hold for the
stochastic Langevin-Kramers evolution with additive noise. His result was
extended to more general diffusion processes by Lebowitz and Spohn in \cite
{LebowSp}. In \cite{Maes}, Maes has traced the origin of \textbf{FR} to the
Gibbsian nature of the statistics of the dynamical histories. Finally, these
relations were put into the language of stochastic thermodynamics by Sekimoto \cite{Sek} and Seifert 
\cite{Sei}. There exist now many reviews on fluctuation relations in the
Markovian context, like \cite{Sch1,Jia1,Liu3} for pure jump process or \cite
{CHCHJAR,Che1,Kurchan2,Jia1,Liu1} for diffusion process, but the extension to
{\bf FR} for general Markov  
process is still under debate \cite{Bau1,Tou1}.

The paper consists of seven Sections and six Appendices. Section 2 sets
the stage and provides notations by briefly stating definitions relevant to 
Markov 
processes. In particular, in Section 2.1, we recall the notions of transition probability,
Markov  
generator, 
stationary state and 
equilibrium
state. In Section 2.2, we introduce the notion of
cotransition probability, cogenerator, current and
velocity operator
. We also elucidate the relation between these objects. Section 3 develops
the kinematics of a Markov  
process \cite{Nel} by defining a set of local
derivatives and 
local velocities associated with such processes. It
is proved in Section 3.2 that these local derivatives appear naturally in
the time derivative of correlation function which appears on the right hand
side of the \textbf{FDT} (\ref{ufdt}). Section 4 investigates the form of
the kinematics elements, 
local velocities and 
velocity
operator, for the three most common  
examples of Markov process which appear
in physics. First is the pure jump process in Section 4.1, which is a process
with no diffusion and deterministic evolution. Second is the
diffusion process, considered in Section 4.2, which is a process that, on
the contrary, neglects the random jumps. Finally, in Section 4.3, we
investigate the less considered case  
which mixes diffusion,
random jumps and deterministic motion  
given by a
stochastic equation with Gaussian and Poissonian white 
noises. The latter noise consists
of a sequence of $\delta $-function shaped pulses with random heights
occurring at randomly distributed times. Such a noise appears in the
physical world, for example, it describes the emission of electrons in
diodes or the counting process of photons. As examples, we study
two physical realizations of such a dynamics involving colloidal particles
trapped on the unit circle. It turns out that analytical computation of the stationary density is possible only for the first realization, and not for the second. Hence, we resort to extensive numerical simulations to obtain the stationary density as well as the  
local velocity for the second realization. 

The first central section which contains novel results
is Section 5 which is devoted to the study of the behavior of a
Markovian system under a perturbation. More precisely, Section 5.1 recalls
the notion of response function to an arbitrary perturbation. Section 5.2
introduces a special family of perturbations, which we call 
Hamiltonian ones  
or generalized Doob $h$-transforms. These include the usual
perturbations considered in the physics literature. 
Section 5.3 proves in a very
simple way, thanks to the language of kinematics elements, that the recent 
\textbf{GFDT} \cite
{Bae1,Bae2,Fal2,Che3,Che4,Gom1,Liu2,Mae3,Pro1,Sei2,SpS0,Spe2} are obtained
in this general Markovian context for the case of a
Hamiltonian perturbation. We also numerically verify the {\bf GFDT} in the context of the example of Section 4.3.2 involving stochastic dynamics with Gaussian and Poissonian white noise. Section 5.4 presents the \textbf{GFDT} which
result from a more general class of perturbations, 
such as a time change \cite{Dem1} or a thermal perturbation pulse 
\cite{Che4,Risk}. The second crucial
section is Section 6. Here, we present global (non-perturbative) versions of
these \textbf{GFDT} which involve  a family of functionals 
called 
exponential martingales in the probability literature \cite{Pal1}. Originally, a 
martingale referred to a class of betting strategies, but this notion 
has now become central to the modern probability theory and characterizes, 
ironically, a model of a fair game. A 
martingale is process whose 
expectation in the future, given the knowledge accumulated up to now, is its
present value \cite{Chu1,Doo1,Fel1,Rev1,Str}. In Section 6.1, we present 
a family of 
exponential martingales, which are natural objects associated with
the Hamiltonian perturbations because they are 
the ratio of the trajectory measures of the perturbed and the 
unperturbed processes. Moreover,
we prove in Section 6.1.3 that they also provide global versions of the 
GFDT. Finally, in Section 6.2, we revisit, in the light of the 
martingale theory,  the usual {\bf FR} for quite general Markov 
processes and underline the relation with the previously 
considered 
exponential martingales.  In particular, this rationalizes the 
typical martingale form $\left\langle \exp
\left( -W\right) \right\rangle =1 $ of the  
Jarzynski equality. Section 7 presents our conclusions.  
The Appendices collect some simple but technical arguments.

\section{ Elements of a Markov  
process}

As mentioned in the introduction, our study deals with 
nonequilibrium systems modeled by Markov  
processes. We begin by
recollecting below some basic properties of a Markov 
process \cite
{App,Blu1,Chu1,Doo1,Fel1,Rev1,Str}. We consider a continuous time Markov
process $\mathbf{x}_{t}$ which takes values in a space $\mathcal{E}$. The space $\mathcal{E}$ could, for
example, be $\mathcal{R}^{d}$ or a counting space.

\subsection{Transition probability, stationary state and equilibrium}

The dynamics of the process is given by a family of transition functions\footnote{Presence of two time indices is a result of the non-homogeneous time character of the process. Such a process is sometimes called nonstationary
in physics. For time-homogeneous process, we define $P_{t-s} \equiv P_{s}^{t} $.} $P_{s}^{t}(x,dy)$ which satisfy the Chapmann-Kolmogorov rule:
\begin{equation}
\int P_{s}^{u}(x,dy)P_{u}^{t}(y,dz)=P_{s}^{t}(x,dz)~~~~~~~~\forall ~~s\leq
u\leq t,  \label{CK}
\end{equation}
where $P_{s}^{u}(x,dy)$ is the probability that the process has the value $\left[y,y+dy\right]$ at time $u$, conditioned on the fact that it had the value $x$ at time $s$.
Here, and in the following, the notation $dy$ represents the Lebesgue measure or
the counting measure, depending on $\mathcal{E}$. We will assume for
simplicity that the transition functions and all other relevant functions
admit a density with respect to this measure (i.e., $
P_{s}^{t}(x,dy)=P_{s}^{t}(x,y)dy)$. Moreover, we consider processes without death
or explosion, i.e., with so-called honest transition probabilities, such
that one has the normalization condition $\int dyP_{s}^{t}(x,y)=1$. This
could be easily achieved in general, e.g., by enlarging the space to include
a coffin state. It will be useful to think of the transition functions as 
linear operators $P_{s}^{t}$ which form an inhomogeneous semi-group, and which are defined by
their action on a bounded function $f$ in $\mathcal{E}$ in the following way: 
\begin{equation}
P_{s}^{t}[f](x)\equiv \int dyP_{s}^{t}(x,y)f(y).
\end{equation}
The family of transition functions of a Markov process which can be written
down explicitly is very restrictive. Hence, it is more practical to
define the generator $L_{t}$ of this inhomogeneous semi-group, under
appropriate regularity conditions \cite{Chu1,Rev1}, by the following
equation: 
\begin{equation}
P_{s}^{t}=\overrightarrow{\exp }\left( \int_{s}^{t}L_{u}\,du\right) \equiv
\sum_{n}\int_{s\leq s_{1}\leq s_{2}\leq \ldots \leq s_{n}\leq
t}\prod_{i=1}^{n}ds_{i}L_{s_{1}}\circ L_{s_{2}}\circ\ldots\circ L_{s_{n}}.  \label{gen}
\end{equation}
This equation is equivalent to the forward and backward Kolmogorov equation, given, respectively, 
by 
\begin{equation}
\partial _{t}P_{s}^{t}=P_{s}^{t}\circ L_{t},\text{ \ \ and \ }\partial
_{s}P_{s}^{t}=-L_{s}\circ P_{s}^{t}.  \label{FK}
\end{equation}
Here, the symbol $\circ $ means composition of operators. Also, the initial condition is $P_{s}^{s}=\mathcal{I}$. For the transition function to be honest, the
generator must obey $L_{t}[1]=0$, where $1$ is the function which is equal to $
1 $ on $\mathcal{E.}$ If the initial measure of the process is $\mu
_{0}(dx)=\rho _{0}(x)dx$, we may define the averages of a functional of the
process $\mathbf{x}$ as 
\begin{equation}
\left\langle F\right\rangle \equiv \int \mu _{0}(dx)\mathbf{E}_{t_{0},x}\left(
F[x]\right) ,
\end{equation}
where $\mathbf{E}_{t_{0},x}$ stands for the expectation of the functional of the process $
\mathbf{x}$ with the initial condition $x_{t_{0}}=x$. Next, it will be
useful to define a path measure $M_{\mu _{0},\left[ s,t\right] }[dx] \equiv dM_{\mu _{0},\left[ s,t\right] }[x]$  on
the space of trajectories by the following equation:
\begin{equation}
\left\langle F\right\rangle \ =\ \int F[x]\,M_{\mu _{0},\left[ s,t\right]
}[dx],
\end{equation}
where $F$ is a functional of the path from time $s$ to time $t$. The
instantaneous (or single time) probability density function (PDF) of the
process is given by 
\begin{equation}
\rho _{t}(x)=\left\langle \delta (x_{t}-x)\right\rangle .  \label{pdf}
\end{equation}
Its time evolution may be deduced from (\ref{FK}). We obtain the following
Fokker-Planck equation:
\begin{equation}
\partial _{t}\rho _{t}=L_{t}^{\dagger }[\rho _{t}],  \label{eqm}
\end{equation}
where $L_{t}^{\dagger }$ is the formal adjoint of $L_{t}$ with respect to
the Lebesgue (or counting) measure. A \textbf{stationary state} ($\rho
_{t}\equiv \rho $) then satisfies the equation 
\begin{equation}
L_{t}^{\dagger }[\rho ]=0.  \label{INV}
\end{equation}
Further, one says that the process is in 
equilibrium, i.e.,  it
satisfies the infinitesimal detailed balance relation if the following condition for the generator is satisfied 
\footnote{
Note that, with this definition, a non-homogeneous process can be in 
equilibrium. We will see examples of diffusion process with this surprising
property in Section \ref{difpro}.}: 
\begin{equation}
\rho \circ L_{t}\circ \rho ^{-1}=L_{t}^{\dagger }.  \label{DB'}
\end{equation}
If the process is time-homogeneous, the above equation is equivalent to the usual detailed balance condition for the transition function:
\begin{equation}
\rho (x)P_{t-s}(x,y)=\rho (y)P_{t-s}(y,x).  \label{DB}
\end{equation}
It will be useful to define two particular families of non-stationary
states. First, one defines the so-called accompanying density $\pi _{t}$
which satisfies the instantaneous relation \cite{Han,Jarz2} 
\begin{equation}
\ L_{t}^{\dagger }[\pi _{t}]=0.  \label{AD}
\end{equation}
Next, we introduce the subclass of accompanying density, that we assume
to be in local detailed balance, such that the generator verifies the
instantaneous time-dependent version of the relation (\ref{DB'}): 
\begin{equation}
\pi _{t}\circ L_{t}\circ \pi _{t}^{-1}=L_{t}^{\dagger }.  \label{LDB}
\end{equation}

\subsection{ Cotransition probability, current and velocity operator.}
\label{Cot}
The two-point density $\left\langle \delta \left( x_{s}-x\right) \delta
\left( x_{t}-y\right) \right\rangle $ of a Markov process is usually
expressed by conditioning with respect to the earlier time $s$, as 
\begin{equation}
\left\langle \delta \left( x_{s}-x\right) \delta \left( x_{t}-y\right)
\right\rangle =\rho _{s}(x)P_{s}^{t}(x,y).  \label{2pf}
\end{equation}
It can also be expressed by conditioning with respect to the later time $t$
in terms of the so-called cotransition probability $P_{s}^{\ast t}$\ \cite
{Dyn1} (sometimes called the backward transition probability \cite{Dyn2,Nel2}
\footnote{
We will not employ this terminology because in our language, the backward process needs also a reversal of time \cite{Che1}.}) as 
\begin{equation}
\left\langle \delta \left( x_{s}-x\right) \delta \left( x_{t}-y\right)
\right\rangle =P_{s}^{\ast t}(x,y)\rho _{t}(y).  \label{2pb}
\end{equation}
This cotransition probability satisfies the Chapman-Kolmogorov equation (\ref
{CK}), but the normalization condition becomes $\int dxP_{s}^{\ast
t}(x,y)=1. $ The relation between the transition and the cotransition probability can
then be expressed by the operator formula $P_{s}^{\ast t}=\rho _{s} \circ P_{s}^{t} \circ \rho _{t}^{-1}$, 
which implies the forward equation\footnote{Here, the density $\rho$ is regarded as a multiplication operator. In the following, depending on the context, we will consider $\rho$ as a function or as an operator.}  
\begin{equation}
\partial _{t}P_{s}^{\ast t}=P_{s}^{\ast t}\left( \rho _{t}\circ L_{t}\circ
\rho _{t}^{-1}-\rho _{t}^{-1}\left( \partial _{t}\rho _{t}\right) \right) .
\end{equation}
We will now introduce a family of operators $L_{t}^{\ast }$, which we call
cogenerators, by the following equation: 
\begin{equation}
L_{t}^{\ast }=\rho _{t}^{-1}\circ L_{t}^{\dagger }\circ \rho _{t}-\rho
_{t}^{-1}\left( \partial _{t}.\rho _{t}\right) \mathcal{I=}\rho
_{t}^{-1}\circ L_{t}^{\dagger }\circ \rho _{t}-\rho _{t}^{-1}L_{t}^{\dagger
}[\rho _{t}]\mathcal{I},  \label{Lb}
\end{equation}
where $\mathcal{I}$ is the identity kernel, so that the cotransition
probability now takes the operatorial form 
\begin{equation}
P_{s}^{\ast t}\equiv \overrightarrow{\exp }\left( \int_{s}^{t}du\left(
L_{u}^{\ast }\right) ^{\dagger }\right) .  \label{sgb-Lb}
\end{equation}
Then, the property $\int dxP_{s}^{\ast t}(x,y)=1$ is equivalent, as before, to 
$L_{t}^{\ast }[1]=0$. For a stationary process (\ref{INV}
), the cogenerator takes the form $L_{t}^{\ast }=\rho ^{-1}\circ
L_{t}^{\dagger }\circ \rho $, which is the adjoint of $L_t$ with respect to the scalar
product with weight $\rho $. It is also interesting to associate a 
\textbf{current
operator} and a \textbf{velocity operator} (which depend  
\ on the initial 
density) with the density $\rho _{t}$ by the
following equations: 
\begin{equation}
J_{t}\equiv \rho _{t}\circ L_{t}-L_{t}^{\dagger }\circ \rho _{t}\text{ \ and
\ }V_{t}\equiv L_{t}-\rho _{t}^{-1}\circ L_{t}^{\dagger }\circ \rho _{t}.
\label{JV}
\end{equation}
The Fokker-Planck equation (\ref{eqm}) can be expressed as 
\begin{equation}
\partial _{t}\rho _{t}+J_{t}[1]=0\text{, \ or, equivalently, \ }\partial
_{t}\rho _{t}+\rho _{t}V_{t}[1]=0.  \label{eqmJV}
\end{equation}
The condition (\ref{INV}) for the density $\rho $ to be stationary can
then be expressed as 
\begin{equation}
J_{t}[1]=0\text{, \ or, equivalently, \ }V_{t}[1]=0.  \label{SS}
\end{equation}
Otherwise, the 
equilibrium condition (\ref{DB'}) becomes
\begin{equation}
J_{t}=0\text{, \ or, equivalently, \ }V_{t}=0.
\end{equation}
Finally, using (\ref{eqm},\ref{Lb},\ref{JV}), we can express the cogenerator
in terms of the 
velocity operator as 
\begin{equation}
L_{t}^{\ast }=L_{t}-V_{t}+V_{t}[1]\mathcal{I.}  \label{Lstardp}
\end{equation}
Then, by (\ref{Lstardp}), 
equilibrium implies $L_{t}^{\ast }=L_{t}$
. 
The converse of this statement is true because the condition $L_{t}^{\ast }=L_{t}$ implies that for any function $f$, one has 
$L_{t}^{\dagger}\left[\rho_{t}f\right]-fL_{t}^{\dagger}\left[\rho_{t}\right]=\rho_{t}L_{t}\left[f\right]$. Then, on integrating by parts over all space, we get \\
$\int dx\left(-2f(x)L_{t}^{\dagger}\left[\rho_{t}\right](x)\right)=0$, which implies that  
 $L_{t}^{\dagger}\left[\rho_{t}\right]=0$, and then $\rho_{t}=\rho$. Finally, the condition $L_{t}^{\ast }=L_{t}$ can be rewritten as the equilibrium condition.

\begin{figure}[th]
\begin{center}
\includegraphics[width=40mm]{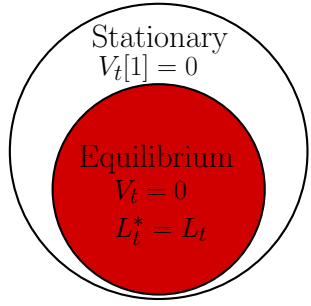}
\end{center}
\caption{The figure illustrates the relation between stationarity,
equilibrium and the condition $L_{t}^{\ast }=L_{t}$, as discussed in the
text.}
\label{progen}
\end{figure}

Figure \ref{progen} illustrates these relations between stationarity, 
equilibrium and the condition of equality between the generator and the
cogenerator. 

\section{ Kinematics of a Markov 
process}

The notion of the 
velocity operator (\ref{JV}) introduced in the last section is quite different from the usual notion of velocity as the derivative of the position. Assume that we want to describe the ``naive'' kinematics of a general Markov  
process. The first difficulty is that the trajectories in general are non-differentiable (as in a diffusion process) or, worse, discontinuous (as in a jump process). This does not allow for a straightforward definition of a velocity. In the sixties, Nelson circumvented this difficulty by introducing the notion of forward and backward \textbf{stochastic derivatives} in his seminal work concerning diffusion process with additive noise \cite{Nel}. Here, we will reproduce the definition of Nelson for a general Markov  
process. In the following, we assume existence conditions on various quantities, with the expectation that these conditions have already been, or, can be established by rigorous mathematical studies.

\subsection{Stochastic derivatives, local velocity}

According to Nelson, a Markov  
process is said to be mean-forward differentiable if the limit $\frac{\lim_{h\rightarrow 0}\left\langle \frac{x_{t+h}-x_{t}}{h}\delta (x_{t}-x)\right\rangle }{\left\langle \delta(x_{t}-x)\right\rangle }$ exists. In this case, this ratio defines the local forward velocity for a process conditioned to be in $x$ at time $t$:  
\begin{equation}
v_{t}^{+}(x)\equiv \frac{\lim_{h\rightarrow 0}\left\langle \frac{
x_{t+h}-x_{t}}{h}\delta (x_{t}-x)\right\rangle }{\left\langle \delta
(x_{t}-x)\right\rangle }.  \label{vf}
\end{equation}
Similarly, the local backward velocity is defined as 
\begin{equation}
v_{t}^{-}(x)\equiv \frac{\lim_{h\rightarrow 0}\left\langle \frac{
x_{t}-x_{t-h}}{h}\delta (x_{t}-x)\right\rangle }{\left\langle \delta
(x_{t}-x)\right\rangle }.  \label{vb}
\end{equation}
The 
local symmetric velocity and the local osmotic velocity are
defined as 
\begin{equation}
v_{t}(x)\equiv \frac{v_{t}^{+}(x)+v_{t}^{-}(x)}{2}\text{ \ and \ }
o_{t}(x)\equiv \frac{v_{t}^{+}(x)-v_{t}^{-}(x)}{2}.  \label{sov}
\end{equation}
In the same spirit, he defined the 
stochastic forward, backward and
symmetric 
derivatives of function $f_{t}(x_{t})$ of the process
as 
\begin{equation}
\left\{ 
\begin{array}{c}
\frac{d_{+}f}{dt}(x)\equiv \frac{\lim_{h\rightarrow 0}\left\langle \frac{
f(t+h,x_{t+h})-f(t,x_{t})}{h}\delta (x_{t}-x)\right\rangle }{\left\langle
\delta (x_{t}-x)\right\rangle }, \\ 
\frac{d_{-}f}{dt}(x)\equiv \frac{\lim_{h\rightarrow 0}\left\langle \frac{
f(t,x_{t})-f(t-h,x_{t-h})}{h}\delta (x_{t}-x)\right\rangle }{\left\langle
\delta (x_{t}-x)\right\rangle }, \\ 
\frac{df}{dt}(x)\equiv \frac{\frac{d_{+}f}{dt}(x)+\frac{d_{-}f}{dt}(x)}{2}.
\end{array}
\right.  \label{d}
\end{equation}
Note that the set of forward, backward and symmetric local velocities are
just special cases of derivatives of the function $f_{t}(x_{t})=x_{t}.$ With
the definition of the forward transition probability and the cotransition
probability given in (\ref{2pf}) and (\ref{2pb}), we can rewrite the above
equations as 
\begin{equation}
\left\{ 
\begin{array}{c}
\frac{d_{+}f}{dt}(x)=\lim_{h\rightarrow 0}\frac{1}{h}\int
dyP_{t}^{t+h}(x,y)\left( f(t+h,y)-f(t,x)\right), \\ 
\frac{d_{-}f}{dt}(x)=\lim_{h\rightarrow 0}\frac{1}{h}\int dyP_{t-h}^{\ast
t}(y;x)\left( f(t,x)-f(t-h,y)\right) .
\end{array}
\right.
\end{equation}
A Taylor expansion of these transition probabilities using (\ref{gen}) and (
\ref{sgb-Lb}) gives 
\begin{equation}
\frac{d_{+}}{dt}=\partial _{t}+L_{t}\text{ \ \ and \ \ \ }\frac{d_{-}}{dt}
=\partial _{t}-L_{t}^{\ast }.
\label{d+d-}
\end{equation}
Also, the 
stochastic symmetric derivative becomes 
\begin{equation}
\frac{d}{dt}=\partial _{t}+\frac{L_{t}-L_{t}^{\ast }}{2}.  \label{SD}
\end{equation}
The expression of the cogenerator from (\ref{Lstardp}) allows us to express
the 
stochastic symmetric derivative in (\ref{SD}) in terms of the 
velocity operator (\ref{JV}) as 
\begin{equation}
\frac{d}{dt}=\partial _{t}+\frac{V_{t}-V_{t}[1]\mathcal{I}}{2}.  \label{sd}
\end{equation}
Then, for a steady state (\ref{SS}), $\frac{d}{dt}=\partial _{t}+\frac{V_{t}}{2}$.

We can then deduce that, in the 
equilibrium case, the 
stochastic symmetric derivative takes the form of the partial time
derivative $\frac{d}{dt}=\partial _{t}$, which gives zero while acting on
observables which do not depend explicitly on time. 
The 
local symmetric velocity
, given in (\ref{sov}), now reads 
\begin{equation}
v_{t}^{i}(x)=\frac{L_{t}[x]-L_{t}^{\ast }[x]}{2}=\frac{
V_{t}[x^{i}]-V_{t}[1]x^{i}}{2},  \label{v}
\end{equation}
and then, for a steady state, $v_{t}^{i}(x)=\frac{V_{t}[x^{i}]}{2}$.

\begin{figure}[th]
\begin{center}
\includegraphics[width=80mm]{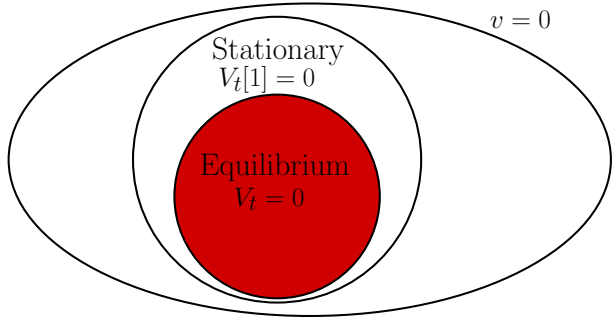}
\end{center}
\caption{The figure illustrates the relation between stationarity,
equilibrium, and vanishing of the 
local symmetric velocity, as discussed in the text.}
\label{fig2}
\end{figure}

It is important to remark that 
equilibrium ($V=0$) implies
vanishing of the 
local symmetric velocity but the converse of this
statement is not true. Figure \ref{fig2} illustrates the relation
between stationarity, 
equilibrium and vanishing of the 
local symmetric velocity.

One of the authors of the present article proved in \cite{Che4} that a diffusion process in the Lagrangian
frame of its mean 
local symmetric velocity takes an 
equilibrium
form, and then the concept of 
equilibrium and 
nonequilibrium become closer
than usually perceived. However, this property is no longer true for a
general process due to the inequivalence between 
equilibrium and
vanishing of the 
local symmetric velocity.

\subsection{Time derivative of two-point correlations}

Here we provide useful formulae for the time derivative of the two-time
($s\leq t$) correlation of observables $U$ and $V$ in terms of the correlation
of stochastic derivatives (forward or backward) of these observables. The two-point
correlation can be expressed in term of the forward transition probability
and cotransition probability, (\ref{2pf}), (\ref{2pb}), as 
\begin{equation}
\left\langle U_{s}(x_{s})V_{t}(x_{t})\right\rangle =\int dxdyU_{s}(x)\rho
_{s}(x)P_{s}^{t}(x,y)V_{t}(y)=\int dxdyU_{s}(x)P_{s}^{\ast t}(x,y)\rho
_{t}(y)V_{t}(y).  \label{coro}
\end{equation}
We then obtain the formula

\begin{equation}
\partial _{t}\left\langle U_{s}(x_{s})V_{t}(x_{t})\right\rangle
=\left\langle U_{s}(x_{s})\frac{d_{+}V_{t}}{dt}(x_{t})\right\rangle \text{
and }\partial _{s}\left\langle U_{s}(x_{s})V_{t}(x_{t})\right\rangle
=\left\langle \frac{d_{-}U_{s}}{ds}(x_{s})V_{t}(x_{t})\right\rangle .
\label{dercor}
\end{equation}
The proofs are direct consequence of the definition of transition and
cotransition probabilities (\ref{gen},\ref{sgb-Lb}) and of forward and
backward stochastic derivatives, and are given in Appendix (\ref{dcor}).
These relations provide motivation for a proof of generalizations of \textbf{
FDT} by involving the 
stochastic derivatives, as discussed later in the paper.

\section{Examples of Markov  
processes}

We will now investigate the form of the 
velocity operator (\ref{JV}
) and of the 
local symmetric velocity (\ref{v}) for the three most
popular examples of Markov  
processes, namely, the pure jump process, the diffusion
process and a process generated by a stochastic equation with 
both Gaussian and Poissonian white noise.

\subsection{Pure jump process}

\label{DP}Roughly speaking, a Markov process is called a pure jump process
(or, a pure discontinuous process) if, after ``arriving'' into a state, the
system stays there for an exponentially-distributed random time interval. It then jumps into another
state chosen randomly, where it spends a random time, and so on. More
precisely, $\mathbf{x}_{t}$ is a pure jump process if, during an
arbitrary time interval $\left[ t,t+dt\right]$, the probability that the
process undergoes one unique change of state (respectively, more that one
change of state) is proportional to $dt$ (respectively, infinitesimal with respect to $
dt)$ \cite{Fel1}. In a countable space, one can show that all Markov
processes (with right continuous trajectories) are of this type, a property
which is not true in a general space. It is usual to introduce the intensity
function $\lambda _{t}(x)$ such that $\lambda _{t}(x)dt+o(\lambda _{t}(x)dt)$ is the
probability that $\mathbf{x}_{t}$ undergoes a random change in the time
interval $\left[ t,t+dt\right] $ if the actual state is $x_{t}=x$. If this
change occurs, then $x(t+dt)$ is distributed with the transition matrix $
T_{t}(x,dy)$. Such a process naturally generalizes a Markov chain to
continuous time. 

We introduce the transition rate of the jump process, which gives the rate
at time $t$ for the transition $x\rightarrow y,$ through 
\begin{equation}
W_{t}(x,dy)\equiv \lambda _{t}(x)T_{t}(x,dy).
\end{equation}
One can prove that, with regularity condition \cite{Fel1,Rev1}, such a
process possesses the generator 
\begin{equation}
L_{t}(x,y)=W_{t}(x,y)-\delta (x-y)\left( \int dzW_{t}(x,z)\right). 
\label{genj}
\end{equation}
The current and the 
velocity operator, given in (\ref{JV}), take
the form of the kernel 
\begin{equation}
J_{t}(x,y)=\rho _{t}(x)W_{t}(x,y)-W_{t}(y,x)\rho _{t}(y)\text{\ and }
V_{t}(x,y)=W_{t}(x,y)-\rho _{t}^{-1}(x)W_{t}(y,x)\rho _{t}(y).  \label{KDP}
\end{equation}
Otherwise, the 
local symmetric velocity (\ref{v}) takes the form 
\begin{equation}
v_{t}(x)=\frac{1}{2}\int V_{t}(x,y)(y-x)dy.
\end{equation}

\subsection{Diffusions processes}

\label{difpro} Here we are interested in a Markov process which has continuous
trajectories. More concretely, the main objects of our study are the
non-autonomous stochastic processes $\,\mathbf{x}_{t}\,$ in $\mathcal{R}^{d}$
(or, more generally, on a $d$-dimensional manifold), described by the
differential equation 
\begin{equation}
\dot{x}\ =\ u_{t}(x)\,+\,\eta _{t}(x)\,,  \label{E0}
\end{equation}
where $\,\dot{x}\equiv \frac{dx}{dt}$, $\,u_{t}(x)\,$ is a time-dependent
deterministic vector field (a drift), and $\,\eta _{t}(x)\,$ is a Gaussian
random vector field with mean zero and covariance 
\begin{equation}
\big\langle\eta _{t}^{i}(x)\,\eta _{s}^{j}(y)\big\rangle\ =\ \delta
(t-s)\,D_{t}^{ij}(x,y)\,.  \label{dep2}
\end{equation}
Due to the white-noise nature of the temporal dependence of $\,\eta_{t}\,$
(typical $\,\eta_{t}\,$ are distributional in time), (\ref{E0}) is a stochastic
differential equation (SDE). We shall consider it with the Stratonovich
convention \cite{Str}, keeping for the Stratonovich SDE's the notation of the
ordinary differential equations (ODE's). The explicit form of generator $
\,L_{t}\,$ which acts on a function $f$ is
\begin{equation}
L_{t}[f]\,=\widehat{u}_{t}^{i}\partial _{i}f+\frac{_{1}}{^{2}}\partial _{j}
\left[ d_{t}^{ij}\partial _{i}f\right] \,,  \label{gend}
\end{equation}
where 
\begin{equation}
d_{t}^{ij}(x)=D_{t}^{ij}(x,x)\qquad \mathrm{and}\qquad \widehat{u}
_{t}^{i}(x)=u_{t}^{i}(x)-\frac{_{1}}{^{2}}\partial
_{y^{j}}D_{t}^{ij}(x,y)|_{y=x}.
\end{equation}
Here, $\widehat{u}_{t}^{i}(x)$ is called the modified drift. A particular
form of \thinspace (\ref{E0}) which is very popular in physics is the
so-called overdamped Langevin form (with the Einstein relation): 
\begin{equation}
\dot{x}^{i}\ =\ -\Gamma _{t}^{ij}(x)\partial _{j}H_{t}(x)\,+G_{t}^{i}(x)+
\frac{1}{2}\partial _{y^{j}}\left. D_{t}^{ij}(x,y)\right| _{x=y}+\,\eta
_{t}^{i}(x)\,\text{ and }d_{t}^{ij}(x)=\frac{2}{\beta }\Gamma _{t}^{ij}(x)
\text{ },  \label{sys}
\end{equation}
where $H_{t}(x)$ is the Hamiltonian of the system (the time index
corresponds to an explicit time dependence), $\Gamma _{t}(x)\,$is a family
of non-negative matrices, \ $G_{t}(x)$ is an external force (or a shear), $
\beta $ the reciprocal of the bath temperature and $\partial _{y^{j}}\left.
D_{t}^{ij}(x,y)\right| _{x=y}$ is an additional spurious term which comes
from the $x$ dependence of the noise. This additional term is chosen in such
a way that the accompanying density (\ref{AD}) is the Gibbs density $\frac{
\exp (-\beta H_{t})}{Z_{t}}$, in the case where the external force is zero ($G=0$). Then, in the case of stationary Hamiltonian
and temperature (i.e., $H_{t}=H,\beta _{t}=\beta $) and without the external
force (i.e, $G=0$), the Gibbs density $\frac{\exp (-\beta H)}{Z}$ is an 
equilibrium density, see (\ref{DB'}). Note that this last case, in the situation where
the matrix $\Gamma _{t}$ depends explicitly on time, is an example of 
a non-homogeneous process in 
equilibrium in the state $\exp (-\beta H).$
The presence of the spurious term $\partial _{y^{j}}\left.
D_{t}^{ij}(x,y)\right| _{x=y}$ was extensively studied in the literature of non-linear Brownian motion \cite{Kil1} and we can see that it vanishes in the
case of linear Brownian motion where $\Gamma _{t}(x)=\Gamma _{t}$. The
overdamped property comes from neglect of the Hamiltonian forces\footnote{
The Fluctuation-Dissipation Theorem with such Hamiltonian force has been
studied in details in \cite{Che3,Che4}.}. In addition to the operator
current, the 
operator velocity (\ref{JV}) and the
local symmetric velocity (\ref{v}), it is usual for this type of process to
introduce the hydrodynamic probability current $j_{t}$, respectively, 
the \textbf{
hydrodynamic velocity} $\widetilde{v}_{t}$, associated with the PDF $\rho
_{t}$, (\ref{pdf}), through 
\begin{equation}
j_{t}=\widehat{u}_{t}\rho _{t}-\frac{d_{t}}{2}(\nabla \rho _{t})\text{ and }
\widetilde{v}_{t}=\widehat{u}_{t}-\frac{d_{t}}{2}(\nabla \ln \rho _{t}),
\label{jv}
\end{equation}
such that the Fokker-Planck equation (\ref{eqm}) takes the form of the
continuity equation, respectively, the hydrodynamical advection equation, 
\begin{equation}
\partial _{t}\rho _{t}+\nabla _{i}j_{t}^{i}=0\text{ \ \ and \ \ }\partial
_{t}\rho _{t}+\nabla _{i}\left( \rho _{t}\widetilde{v}_{t}^{i}\right) =0.
\label{ce}
\end{equation}
A direct calculation, given in Appendix (\ref{cogpd}), shows that
the explicit form of the cogenerator (\ref{Lb}) for a diffusion process is 
\begin{equation}
L_{t}^{\ast }[f]=L_{t}[f]-2\widetilde{v}_{t}.\nabla f,  \label{Lstardiff}
\end{equation}
and we can deduce the form of the 
operator velocity, (\ref{JV}), as 
\begin{eqnarray}
V_{t}[f]&=&\left( L_{t}-\rho _{t}^{-1}\circ L_{t}^{\dagger }\circ \rho
_{t}\right) [f]\nonumber \\
&=&\left( L_{t}-L_{t}^{\ast }-\rho _{t}^{-1}\left( \partial
_{t}.\rho _{t}\right) \mathcal{I}\right) [f]=2\widetilde{v}_{t}.\nabla
f+\left( \rho _{t}^{-1}\nabla _{i}\left( \rho _{t}\widetilde{v}
_{t}^{i}\right) \right) f.  
\label{Ivv}
\end{eqnarray}
Moreover, for a diffusion process, (\ref{Lstardiff}) allows us to
obtain the following hydrodynamical form for the 
stochastic
symmetric derivative and the 
local symmetric velocity. 
\begin{equation}
\frac{d}{dt}=\partial _{t}+\widetilde{v}_{t}.\nabla \text{ \ and \ }v_{t}(x)=
\widetilde{v}_{t}(x).  \label{dspd}
\end{equation}
It then follows that the 
local symmetric velocity is
identical to the 
hydrodynamic velocity, and moreover, with (\ref{Ivv}) , that the 
equilibrium condition ($V_{t}=0)$ is equivalent to the
condition of vanishing of the 
hydrodynamic velocity $\widetilde{v}
_{t}$ or the 
local symmetric velocity in $\mathcal{E}$\footnote{
In \cite{Dar1}, a result in a similar spirit was shown for the characterization
of diffusion processes with additive covariance $d_{t}^{ij}(x)=d\delta
^{ij}$ which possesses a (possibly time-dependent) gradient drift. The
characterization can be written in terms of a second-order 
stochastic
derivative as $\frac{d_{+}d_{+}x_{t}}{dt^{2}}=-\frac{d_{-}d_{-}x_{t}}{dt^{2}}$.}. Also, the form of the drift of an 
equilibrium diffusion is
then 
\begin{equation}
\widehat{u}_{t}=\frac{d_{t}}{2}(\nabla \ln \rho ).
\end{equation}
The link between stationarity, equilibrium and vanishing of local symmetric velocity for a diffusion process is depicted in Fig. 3.

\begin{figure}[th]
\begin{center}
\includegraphics[width=40mm]{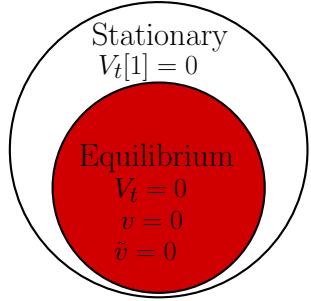}
\end{center}
\par
\label{fig4seehere}
\caption{The set of processes inside the domain marked in red are such that
the local symmetric velocity vanish, but it is also the set of equilibrium jump processes and the set
of processes with vanishing hydrodynamic velocity.}
\end{figure}

\subsection{Stochastic equation with Gaussian and Poissonian white noise}

\label{JD} We now consider a Markov process in continuous space (e.g., $
\mathcal{R}^{d}$) which includes the processes in the last two sections in
the sense that both diffusion and jump can occur. Such processes are very
popular in finance \cite{Con,Mer1}. They are much less popular in physics,
where, after its first study in the beginning of eighties \cite{Han2,Van2},
they were used, for example, to study mechanism of noise-induced transitions 
\cite{San1} or noise-driven transport \cite{Cze1,Luc1}. We consider
processes that are right continuous with a left limit (i.e., ``cadlag'' processes), and we
define $x_{t^{-}}=\lim_{s\uparrow t}x_{s\text{ }}$ and the jump as 
\begin{equation}
\Delta x_{t}=x_{t}-x_{t^{-}}.
\end{equation}
We want to consider a process which follows the evolution of a diffusion
process (\ref{E0}) for most of the time, excepting that it jumps
occasionally, the occurrence of the jump being given by a non-autonomous
Poisson process. More precisely, we will construct such processes by adding
a state-dependent Poisson noise \cite{Por1} to the stochastic differential
equation (\ref{E0}), as 
\begin{equation}
\dot{x}_{t}\ =\ u_{t}(x_{t})\,+\,\eta _{t}(x_{t})+w_{t}(x_{t^{-}})\,,
\label{GPN}
\end{equation}
where, as before, $\eta _{t}(x)\,$ is a Gaussian random vector field (with
Stratonovich convention \cite{Str}) which has mean zero and covariance (\ref
{dep2}). On the other hand, $w_{t}(x)\,$\ is a state-dependent Poisson noise
(that depends on the state $x_{t}$), and is given by 
\begin{equation}
w_{t}(x)=\sum_{i=1}^{N_{t}}y_{i}(x)\delta (t-T_{i}).
\end{equation}
The time $T_{i}$ at which the instantaneous jump occurs are the arrival
times of a non-homogeneous and non-autonomous Poisson process $N_{t}$ with
intensity $\lambda _{t}(x).$ The jump magnitude $y_{i}$ are
mutually-independent random variables, independent of the Poisson process,
and are described by the probability function $b_{t,x}(y)$. This function
gives the probability for a jump of magnitude $y$ while starting from $x$ at
time $t$. Physically, addition of the Poisson noise mimics large
instantaneous inflows or outflows (``big impact'') at the microscopic level.
We remark that this noise contains almost surely a finite number of jumps in every
interval ($\lambda _{t}(x)$ is finite). It is possible to consider a more
general noise, the so-called Levy noise, where this condition is relaxed\footnote{
The process $\mathbf{x}_{t}$ then describes a fairly large class of Markov processes
(of Feller-type) which are governed by Levy-Ito generators 
which acts
on a function $f$ as the integro-differential 
operators \cite{Jac1,Str2} 
\begin{equation}
L_{t}[f](x)\,=\widehat{u}_{t}^{i}(x)\partial _{x^{i}}f+\frac{_{1}}{^{2}}
\partial _{x^{j}}\left[ d_{t}^{ij}(x)\partial _{x^{i}}f\right] \,+\int_{
\mathcal{R}^{d}-\{0\}}\left( f(x+y)-f(x)-\frac{\left( y.\nabla f\right) (x)}{
1+\left| y\right| ^{2}}\right) \nu _{t,x}(dy),
\end{equation}
with the so-called Levy jump measure $\nu _{t,x}(dy)$ which can be infinite 
but is such that, for all $x$ and $t$, the
condition $\int_{\mathcal{R}^{d}-\{0\}}\frac{\left| y\right| ^{2}}{1+\left|
y\right| ^{2}}\nu _{t,x}(dy)<\infty $ is verified.}. The mathematical theory
of general stochastic differential equation with a Levy noise and the theory
of stochastic integration with respect to a (possibly discontinuous) more general  
(semi-
martingale) noise are well established \cite{App,Bas1}. In the present case, we
will just use from this theory the form of the Markov 
generator which,
for the process (\ref{GPN}), is an integro-differential operator, given by 
\begin{equation}
L_{t}=L_{t}^{D}+L_{t}^{J}.  \label{gjd}
\end{equation}
Here, the diffusive part $L_{t}^{D}$ is given by (\ref{gend})
and the jump part $L_{t}^{J}$ is given by (\ref{genj}), with 
\begin{equation}
W_{t}(x,y)=\lambda _{t}(x)b_{t,x}(y-x).  \label{TRP}
\end{equation}
The class of process (\ref{GPN}) possesses some famous particular cases.

\begin{itemize}
\item  The piecewise deterministic process \cite{Dav1} is the case where
there is no Gaussian noise ($\eta _{t}(x)=0).$ Then $x_{t}$ follows a
deterministic trajectory interrupted by jumps of random timing and
amplitudes.

\item  The interlacing Levy Processes \cite{App} is the case where the drift
is constant and homogeneous ($u_{t}(x_{t})=u$), the Gaussian noise is
additive and stationary ($d_{t}^{ij}(x)=d^{ij})$, and the Poisson white
noise is state-independent and stationary ($\lambda _{t}(x_{t})=\lambda $
and $b_{t,x}(y)=b(y)).$ This process belongs to the class of Levy process 
\cite{App}, with independent and homogeneous increments.
\end{itemize}

We will now investigate the form of the kinematics elements: the 
velocity operator, (\ref{JV}), and the 
local symmetric velocity, (
\ref{v}). Similar to (\ref{gjd}), these two objects can be split into a
diffusive part and a jump part such that 
\begin{equation}
V_{t}=V_{t}^{D}+V_{t}^{J}\text{ \ \ and \ }v_{t}=v_{t}^{D}+v_{t}^{J}.
\label{cejd}
\end{equation}
On using (\ref{a1}), we can express the diffusive part $V_{t}^{D}$ as (\ref
{Ivv}) 
\begin{eqnarray}
V_{t}^{D}&\equiv& L_{t}^{D}-\rho _{t}^{-1}\circ L_{t}^{D,\dagger }\circ \rho
_{t}\text{ } \nonumber \\
&=&\rho _{t}^{-1}\nabla _{i}\left( \rho _{t}\left( \widehat{u}
_{t}^{i}-\frac{d_{t}^{ij}}{2}\left( \nabla _{j}\ln \rho _{t}\right) \right)
\right) +2\left( \widehat{u}_{t}^{i}-\frac{d_{t}^{ij}}{2}\left( \nabla
_{j}\ln \rho _{t}\right) \right) \nabla _{i}.  \label{vjdd}
\end{eqnarray}
The jump part $V_{t}^{J}$\ is given by (\ref{KDP}) with (\ref
{TRP}). Similarly, the diffusive part of the 
local symmetric velocity reads 
\begin{equation}
v_{t}^{D}(x)=\widehat{u}_{t}^{i}(x)-\frac{d_{t}^{ij}(x)}{2}\left( \nabla
_{j}\ln \rho _{t}(x)\right) ,  \label{vspn1}
\end{equation}
while the jump part of the 
local symmetric velocity reads 
\begin{equation}
v_{t}^{J}(x)=\frac{\int dyW_{t}(x,y)(y-x)-\rho _{t}^{-1}(x)\int dy\rho
_{t}(y)W_{t}(y,x)(y-x)}{2}.  \label{vspn2}
\end{equation}
Finally, the 
stochastic symmetric derivative (\ref{sd}) takes the
form 
\begin{equation}
\frac{d}{dt}=\partial _{t}+\left( \widehat{u}_{t}^{i}-\frac{d_{t}^{ij}}{2}
\left( \nabla _{j}\ln \rho _{t}\right) \right) \nabla _{i}+\frac{W_{t}-\rho
_{t}^{-1}\circ W_{t}^{\dagger }\circ \rho _{t}-\lambda _{t}\mathcal{I}+\rho
_{t}^{-1}W_{t}^{\dagger }[\rho _{t}]\mathcal{I}}{2}.  \label{ddjp}
\end{equation}
Here, we are in the general case where the link between 
equilibrium
($V=0)$ and 
local symmetric velocity is shown in Fig. \ref{fig2}.
However, we remark that the condition 
\begin{equation}
\widehat{u}_{t}=\frac{d_{t}}{2}(\nabla \ln \rho )\text{ \ and \ }\rho
(x)W_{t}(x,y)=\rho (y)W_{t}(y,x)  \label{cs}
\end{equation}
is a sufficient and a necessary condition to be in 
equilibrium (
$V=0)$
\footnote{That the condition is necessary follows from the fact we can split up the kernel $V_{t}$ into a regular and a distributional part, and both should vanish to ensure that $V_{t}=0$.}.

 A particular form of such jump diffusion process (\ref{GPN}), that we
call jump Langevin equation, is obtained from the Langevin equation (\ref
{sys}) by adding a Poisson noise $w_{t}$, as 
\begin{equation}
\dot{x}^{i}\ =\ -\Gamma _{t}^{ij}(x)\partial _{j}H_{t}(x)\,+G_{t}^{i}(x)+
\frac{1}{2}\partial _{y^{j}}\left. D_{t}^{ij}(x,y)\right| _{x=y}+\,\eta
_{t}^{i}(x)+w_{t}^{i}(x),  \label{lepe}
\end{equation}
with $d_{t}^{ij}(x)=\frac{2}{\beta }\Gamma_{t}^{ij}(x)$ such that the transition rate, (\ref{TRP}), takes the particular form
(Kangaroo process \cite{Bri}) 
\begin{equation}
W_{t}(x,y)=r\exp \left( -\frac{\beta }{2}\left( H_{t}(y)-H_{t}(x)\right)
\right),   \label{sftr}
\end{equation}
where $r$ is real. The accompanying density (\ref{AD}), in the case without external
force ($G_{t}=0)$, is the Gibbs density $\frac{\exp (-\beta H_{t})}{Z_{t}}$. If, in addition, we have a stationary Hamiltonian ($H_{t}=H)$, such processes verify the sufficient 
equilibrium condition (\ref{cs}) in
this Gibbs density $\rho (x)=\frac{\exp (-\beta H(x))}{Z}.$ We will now
consider physical examples of jump diffusion process (\ref{GPN}) .

\subsubsection{Example 1 :\ Interlacing Levy process on the unit circle}

The most elementary example of an interlacing Levy process which describes a 
nonequilibrium system is a particle on a unit circle subject to a
constant force $G$, as 
\begin{equation}
\dot{\theta}_{t}=G+\eta _{t}+w_{t},  \label{intlev}
\end{equation}
with an additive and stationary Gaussian white noise ($d_{t}^{ij}(\theta
)=d) $ and a state-independent and stationary Poisson white noise ($\lambda
_{t}(\theta )=\lambda $ and $b_{t,\theta }(\theta ^{\prime })=b(\theta
^{\prime })).$ Moreover, the jump amplitude is a periodic function, $b(\theta
)=b(\theta +2\pi )$. The Fokker-Planck equation (\ref{eqm}) becomes, with (
\ref{gjd}), 
\begin{equation}
\partial _{t}\rho _{t}(\theta )=-G\partial _{\theta }\rho _{t}(\theta )+
\frac{d}{2}\partial _{\theta \theta }^{2}\rho _{t}(\theta )-\lambda \rho
_{t}(\theta )+\lambda \int_{0}^{2\pi }d\theta ^{\prime }b(\theta -\theta
^{\prime })\rho _{t}(\theta ^{\prime }).
\end{equation}
Then, the process possesses an invariant probability distribution with the
constant density $\rho (\theta )=\frac{1}{2\pi }$. This is true also in the
absence of Poisson noise ($\lambda =0)$ or Gaussian noise ($d=0).$ For the
stationary process, where we take the invariant density as initial density,
the 
velocity operator (\ref{cejd},\ref{vjdd}) takes the form 
\begin{equation}
V[f](\theta )=2G\partial _{\theta }f+\lambda \int_{0}^{2\pi }d\theta
^{\prime }\left( b(\theta ^{\prime }-\theta )-b(\theta -\theta ^{\prime
})\right) f(\theta ^{\prime }).  \label{intlevvo}
\end{equation}
In the absence of external force (i.e., $G=0)$, we see that the Poisson
noise transforms an 
equilibrium state to a 
nonequilibrium
steady state if $b$ is not an even function. Finally, the 
local symmetric velocity takes the form (\ref{cejd},\ref{vspn1},\ref{vspn2}) 
\begin{equation}
v(\theta )=G+\frac{\lambda }{2}\int_{0}^{2\pi }d\theta ^{\prime }\theta
^{\prime }\left( b\left( \theta ^{\prime }-\theta \right) -b\left( \theta
-\theta ^{\prime }\right) \right).
\end{equation}
For example, if we choose
the probability of the jump distribution as $b(\theta )=\frac{1+\sin (\theta
)}{2\pi },$ then the 
local symmetric velocity in the steady state
takes the form $
v(\theta )=G+\frac{\lambda }{2\pi }\int_{0}^{2\pi }d\theta ^{\prime }\sin
\left( \theta ^{\prime }-\theta \right) \theta ^{\prime }=G-\lambda \cos
(\theta )$.
So, despite the fact that the Poisson noise does not change the invariant
density, it changes the 
local symmetric velocity which is no
longer constant around the circle. For example, if $G<\lambda$, it includes
regions of the circle where the local transport is in the reverse sense to the
external force.

\subsubsection{Example 2: Jump Langevin equation on the unit circle}

\label{sjlec} 
We consider a particular case of (\ref{lepe}), namely, 
\begin{equation}
\dot{\theta}_{t}=-\partial_{\theta }H+G+\eta _{t}+w_{t},
\label{jlec}
\end{equation}
which describes the angular position of an overdamped particle on a circle. The Hamiltonian $H$ is $2\pi$-periodic, the force $G$ is a constant, the Gaussian white noise $\eta_{t}$ has the covariance $\langle \eta_{s}\eta _{t} \rangle =\frac{2}{\beta }\delta (t-s)$, and the transition rates of the state-dependent Poisson white noise are given by (\ref{sftr}). Such systems without the Poisson noise ($r=0$) have been realized with a colloidal particle kept by an optical tweezer on a nearly circular orbit \cite{Gom1}. In these experiments, $H(\theta )=a\sin\theta$. In this case, the invariant density takes the form \cite{Fal2} 
\begin{eqnarray}
\rho (\theta )&&=Z^{-1}\exp(-\beta \{H(\theta)-G\theta\})\nonumber \\
&&\times \Big(\int\limits_{0}^{\theta}\exp(\beta \{H(\vartheta )-G\vartheta\})d\vartheta + \exp (2\pi \beta G)\int\limits_{\theta}^{2\pi}\exp(\beta\{H(\vartheta)-G\vartheta\})d\vartheta \Big), \nonumber \\
\label{rhonP}
\end{eqnarray}
where $Z$ is the normalization factor. The corresponding 
local symmetric velocity (also the 
hydrodynamic velocity in the present context) takes the form 
\begin{equation}
v(\theta )=\ \beta ^{-1}Z^{-1}\frac{\exp \left( 2\pi \beta G\right)-1}{\rho(\theta)}.
\label{vthetanP}
\end{equation}

\begin{figure}[h!]
\begin{center}
\includegraphics[width=100mm]{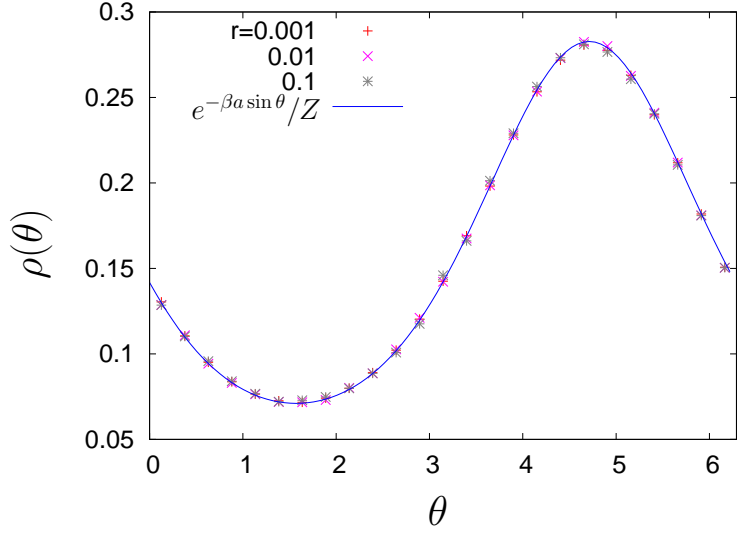}
\end{center}
\caption{The points in the figure show the equilibrium density $\rho(\theta)$, obtained from numerical simulations of the dynamics (\ref{jlec}) with $H=a\sin \theta$, and with $G=0$, $a=0.87s^{-1}$, $\beta=0.8s$, and for three values of $r=0.001,0.01$, and $0.1$. It can be seen that the results do not depend on the value of $r$. In the figure, the results of numerical simulations have also been compared with the analytical result given in (\ref{eeqq}) with $H=a\sin \theta$, and is represented in the figure by the continuous line.}
\label{fig4}
\end{figure}

However, with the Poisson noise ($r\neq 0$), it is not possible to obtain 
analytically the form of the stationary state, except in the 
equilibrium case (i.e., without external force, $G=0$), where the 
equilibrium density is 
\begin{equation}
\rho(\theta)=Z^{-1}\exp(-\beta H(\theta )),
\label{eeqq}
\end{equation}
and the 
local symmetric velocity is zero. We realize a numerical simulation of the system (\ref{jlec}) with $a=0.87s^{-1}$ and $\beta=0.8s$ (these values for $a$ and $\beta$ are close to those used in the experiment \cite{Gom1}), but with a non-vanishing Poisson noise ($r\neq 0$). We can imagine for example that it is once again the laser beam which produces the two noise. We first verify numerically that we find the equilibrium density (\ref{eeqq}) for three values of $r=0.001,0.01$, and $0.1$ in the case $G=0$. The results of the numerical simulation are shown in Fig. \ref{fig4} which confirm the independence of the equilibrium density on the Poisson noise.

\begin{figure}[h!]
\begin{center}
\includegraphics[width=118mm]{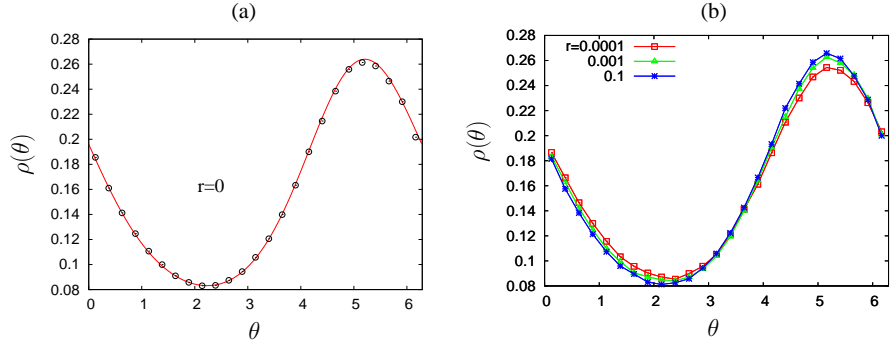}
\end{center}
\caption{(a) The points represent the stationary density $\rho(\theta)$, obtained from numerical simulation of the dynamics (\ref{jlec}) with $H=a \sin \theta$, and with $G=0.85s^{-1}$, $a=0.87s^{-1}$, $\beta=0.8s$, but without the Poisson noise ($r=0$). As expected, the points may be seen to lie on the continuous line representing the exact result in (\ref{rhonP}). (b) Here, we show the stationary density $\rho(\theta)$, obtained from numerical simulations of the dynamics (\ref{jlec}) with $G=0.85s^{-1}$, $a=0.87s^{-1}$, $\beta=0.8s$, and for $r=0.0001,0.001$, and $0.1$. It is easily seen that $\rho(\theta)$ depends on the value of $r$, thereby hinting at the important role played by the Poisson noise.}
\label{fig5}
\end{figure}

\begin{figure}[h!]
\begin{center}
\includegraphics[width=100mm]{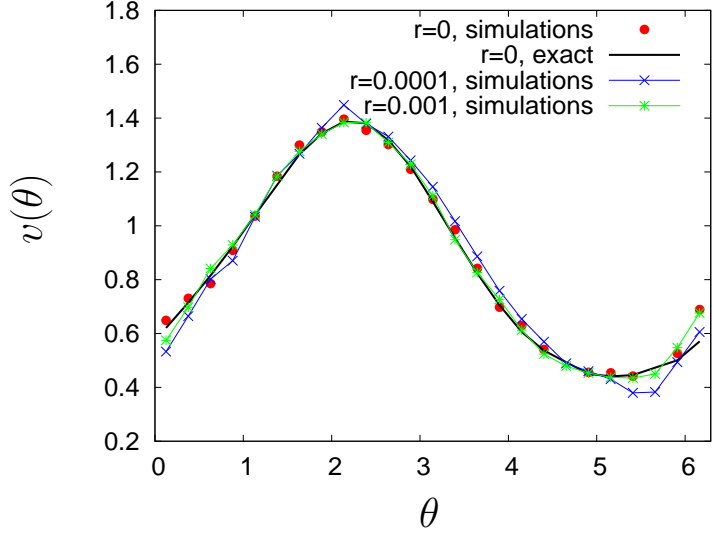}
\end{center}
\caption{(a) The figure shows the local symmetric velocity $v(\theta)$ for the dynamics (\ref{jlec}) with $H=a\sin \theta$, and with $G=0.85s^{-1}$, $a=0.87s^{-1}$, $\beta=0.8s$, and for $r=0$ (no Poisson noise), $0.0001$, and $0.001$. The points are obtained from numerical simulations of the dynamics and use of the formula (\ref{vtheta}). The exact result for the case $r=0$ is given by (\ref{vthetanP}). It is easily seen that $v(\theta)$ depends on the value of $r$, i.e., on the details of the Poisson noise in the dynamics.}
\label{fig6}
\end{figure}

Next, we investigate numerically the case where the external force takes the value of the experiments \cite{Gom1} ($G=0.85s^{-1}$) for three different values of $r$ (which characterizes the role of the Poisson noise), namely, $r=0.0001,0.001$, and $0.1$. The corresponding forms of the stationary state distribution are shown in Fig. \ref{fig5}(b). From the figure, it is evident that in the presence of the external force, the form of the non-equilibrium stationary state depends on $r$, thereby underlying the importance of the Poisson noise. This is to be contrasted with the result for the case depicted in Fig. \ref{fig4}, i.e., with $G=0$, when the form of the equilibrium stationary state is independent of $r$. Corresponding to the non-equilibrium stationary state for $G \ne 0$, the 
local symmetric velocity (\ref{cejd},\ref{vspn1},\ref{vspn2}) is given by 
\begin{eqnarray}
v(\theta ) &=&-a\cos (\theta )+G-\frac{1}{\beta }\partial _{\theta }\left(
\ln \rho (\theta )\right) \nonumber \\
&&+\frac{r\exp (\frac{a\beta }{2}\sin (\theta ))}{2}
\left( \int_{0}^{2\pi }d\theta ^{\prime }\exp \left( -\frac{\beta a}{2}\sin
(\theta ^{\prime })\right) \theta ^{\prime }\right) \nonumber  \\
&&-\frac{r\rho ^{-1}(\theta )\exp (-\frac{a\beta }{2}\sin (\theta ))}{2}
\left( \int_{0}^{2\pi }d\theta ^{\prime }\rho (\theta ^{\prime })\exp \left( 
\frac{\beta a}{2}\sin (\theta ^{\prime })\right) \theta ^{\prime }\right). \nonumber \\
\label{vtheta}
\end{eqnarray}

In Fig. \ref{fig6}, we show the local symmetric velocity $v(\theta)$ for the dynamics (\ref{jlec}) with $G=0.85s^{-1}$, $a=0.87s^{-1}$, $\beta=0.8s$, and for $r=0$ (no Poisson noise), $0.0001$, and $0.001$. It is clear from the figure that the quantity $v(\theta)$ depends on the details of the Poisson noise in the dynamics.

In Section \ref{numericalFDT}, we will use the dynamics (\ref{jlec}) as a model system to verify the {\bf GFDT} by extensive numerical simulations.

\section{Perturbation of a Markov 
process: the Fluctuation-Dissipation Theorem}

Suppose that our dynamics evolves for $t\leq0$ with a given Markovian dynamics of generator $
L_{t}$ and then suddenly, at time $t=0$, we perturb the dynamics such that
the new Markov  
generator $L_{t}^{\prime }$ becomes 
\begin{equation}
L_{t}^{\prime }=L_{t}+k_{t}N_{t},  \label{inf}
\end{equation}
with $k_{t}$ a real function, sometimes called the response field, and $
N_{t} $ an operator. We will assume that the perturbed process still has the
property to have honest transition probability (i.e., $N_{t}[1]=0).$ The 
\textbf{FDT }concerns the relation between correlation functions, (\ref{coro}), in the
unperturbed state and response functions in the case of a
small perturbation (i.e., $k_{t}$ infinitesimal).

\subsection{Response function}

The linear response theory allows to express the variation of the average of an 
observable under the perturbation as 
\begin{equation}
\left. \frac{\delta \left\langle A_{t}(x_{t})\right\rangle ^{\prime }}{
\delta k_{s}}\right| _{k=0}=\left\langle \left( \rho _{s}^{-1}N^{\dagger
}[\rho_{s}]\right) (x_{s})A_{t}(x_{t})\right\rangle ,  \label{rep}
\end{equation}
where $\left\langle {}\right\rangle^{\prime }$ denotes expectation for the
process with the generator $L_t^{\prime }.$ The proof of this relation is
given in Appendix (\ref{linres}) for the convenience of the reader. Note
however that this result is known for a long time in the physics literature 
\cite{Agarw,Han,Kubo0,Risk} and now has a mathematically rigorous
formulation (Definition $2.5$ in \cite{Dem1}). This relation, besides being the basis for the {\bf FDT}, allows to prove the Green-Kubo relation \cite{Kubo2} in the case of homogeneous perturbation ($k_t=k$) of a stationary dynamics. Note that other higher order relations may be derived in the context of the non-linear response theory \cite{Lip2}.

\subsection{Hamiltonian perturbation class or generalized Doob $h$-transform}

\label{hp} We will see that the form of the perturbation is the central
point of the \textbf{FDT}, and it does not make sense to talk of \textbf{FDT}
without giving its form. We want to begin by studying a
class of (non-infinitesimal) perturbation of the Markov process such that the
transformation of the generator can be expressed in terms of a family of
non-homogeneous positive function $h_t$, as 
\begin{equation}
L_{t}^{^{\prime }}=h_{t}^{-1}\circ L_{t}\circ h_{t}-h_{t}^{-1}L_{t}\left[
h_{t}\right] \equiv L_{t}^{h}.  \label{pert}
\end{equation}
In the case where $\frac{d_{+}h_{t}}{dt}=\partial _{t}h_{t}+L_{t}\left[ h_{t}
\right] =0$ ($h_{t}(x)$ is the so-called space time harmonic function), such
a transformation is classical in the probability literature and is called
the Doob $h$-transform (or gauge transformation in physics literature). 
This was first introduced
by Doob (\cite{Doo1}; see also chapter 11 of \cite{Chu1}), and plays an
important role in the potential theory. We remark that if $h_{t}(x)$ is
space-time harmonic, then 
$h_{t}(x_{t})$ is a \textbf{martingale}, i.e., 
\begin{equation}
h_{t}(x)=\mathbf{E}_{t,x}\left( h_{T}(x_{T})\right) ~~~~~~~~~\forall ~~T\geq t.
\label{martingale}
\end{equation}
By introducing the symmetric bilinear operator $\Gamma $ (the so-called
''carre du champs'' \cite{Rev1}, which can be roughly translated into English as ``square of the field'', such that $\Gamma _{t}(f,g)=L_{t}\left[ fg
\right] -fL_{t}\left[ g\right] -L_{t}\left[ f\right] g$, the perturbed
generator can be expressed in the form $L_{t}^{h}=L_{t}+h_{t}^{-1}\Gamma (h_{t},)$. A remarkable property of this type of perturbation appears if we restrict to
a subclass of unperturbed processes which are in so-called local detailed
balance (\ref{LDB}) with the Gibbs density $\pi _{t}=\exp (-\beta H_{t})$.
We then have the relation 
\begin{eqnarray}
h_{t}^{2}\pi _{t}\circ L_{t}^{h}\circ \left( h_{t}^{2}\pi _{t}\right)
^{-1}&&=h_{t}\pi _{t}\circ L_{t}\circ \left( \pi _{t}\right)
^{-1}h_{t}^{-1}-h_{t}^{-1}L_{t}(h_{t})\nonumber \\
&&=h_{t}\circ L_{t}^{\dagger }\circ
h_{t}^{-1}-h_{t}^{-1}L_{t}(h_{t})= \left(L_{t}^{h}\right)^{\dagger },
\end{eqnarray}
which implies that, for the perturbed process, the density, given by 
\begin{equation}
\pi _{t}^{h}=\pi _{t}h_{t}^{2}=\exp (-\beta H_{t}+2\ln  h_{t} ),
\label{dojp}
\end{equation}
is also in local detailed balance. This property of conservation of
instantaneous infinitesimal detailed balance under the perturbation of the
Hamiltonian $H\rightarrow H-\frac{2}{\beta }\ln h_{t}$\ \ is the first
justification for the name ``Hamiltonian perturbation'' that we chose for this type
of perturbation. However, we stress that this perturbation, although called here ``Hamiltonian perturbation", is applicable to general Markov processes which do not have an underlying Hamiltonian which generates the dynamics. Moreover, for a general diffusion process, we can easily calculate (see
Appendix (\ref{pcarc})) the operator ''carre du champs'' 
\begin{equation}
\Gamma _{t}(f,g)=d_{t}^{ij}\left( \nabla _{i}f\right) \left( \nabla
_{j}g\right) .  \label{carc}
\end{equation}
Then the perturbed generator (\ref{pert}) is 
\begin{equation}
L_{t}^{h}=L_{t}+d_{t}^{ij}\nabla _{j}\left( \ln \left| h_{t}\right| \right)
\nabla _{i},  \label{adddrift}
\end{equation}
so that there is just a change of the drift term, $u_{t}^{i}\rightarrow
u_{t}^{i}+d_{t}^{ij}\nabla _{j}(\ln h_{t}).$ In the subcase of an overdamped
Langevin process (\ref{sys}), the perturbed process (\ref{pert}) 
becomes 
\begin{equation}
\dot{x}^{i}\ =\ -\Gamma _{t}^{ij}(x)\partial _{j}\left( H_{t}(x)-\frac{2}{
\beta }\ln  h_{t} \right) +G_{t}^{i}(x)+\frac{\partial
_{y^{j}}\left. D_{t}^{ij}(x,y)\right| _{x=y}}{^{2}}+\,\eta _{t}^{i}(x)\,.
\label{dolan}
\end{equation}
So we see that the perturbation in (\ref{pert}) is once again equivalent to
change of the Hamiltonian, $H\rightarrow H-\frac{2}{\beta }\ln h_{t}$. Now
we want to show that the type of perturbation in (\ref{pert}) includes the
perturbation usually considered in the articles on \textbf{FDT} that exist in the
literature.

\begin{itemize}
\item  For pure jump process, it is usual to ask precisely the property of
conservation of this local detailed balance for the Gibbs density $\pi
_{t}=\exp (-\beta H_{t})$ under the perturbation of the Hamiltonian $
H\rightarrow H-k_{t}O_{t}$. We see from (\ref{dojp}) that this perturbation
of the Hamiltonian is of the type in (\ref{pert}), with the choice 
\begin{equation}
h_{t}=\exp \left( \frac{\beta }{2}k_{t}O_{t}\right).   \label{hjp}
\end{equation}
This implies the following transformation for the transition rates. 
\begin{equation}
W_{t}^{h}(x,y)=\exp \left( -\frac{\beta }{2}k_{t}O_{t}(x)\right)
W_{t}(x,y)\exp \left( \frac{\beta }{2}k_{t}O_{t}(y)\right) .  \label{ppj}
\end{equation}
This is the perturbation considered recently in \cite{Bae1} and earlier in \cite
{Die1} for finding the \textbf{GFDT} in this pure jump process set-up.

\item  For overdamped Langevin process, it is usual \cite
{Han,Kubo0,MPRV,Risk} to do a perturbation of the Hamiltonian, $H\rightarrow
H-k_{t}O_{t}$, in (\ref{sys}). With (\ref{dolan}), we see that it is of the
type in (\ref{pert}) with once again (\ref{hjp}) valid.

\item  Finally, we remark that for a jump diffusion process of type (\ref
{GPN}), this perturbation consists of a change of the drift according to $
u_{t}^{i}\rightarrow u_{t}^{i}+d_{t}^{ij}\nabla _{j}(\ln h_{t})$ and
simultaneously, a perturbation of the jump process by replacing the
transition rates (\ref{TRP}) by 
\begin{equation*}
W_{t}^{h}(x,y)=h_{t}^{-1}(x)W_{t}(x,y)h_{t}(y).
\end{equation*}
For the jump Langevin process (\ref{lepe}), with the transition rates (\ref
{sftr}) for the Poisson noise, we can prove easily that the choice (\ref{hjp})
in (\ref{pert}) is equivalent to the perturbation of the Hamiltonian
according to $H\rightarrow H-k_{t}O_{t}.$
\end{itemize}

\subsection{Fluctuation-Dissipation Theorem for Hamiltonian perturbation}

\label{fdth}In the case of an infinitesimal $h_t$ function, 
\begin{equation}
h_{t}(x)=1+k_{t}B_{t}(x)+O(k^{2}),
\end{equation}
we find that the Hamiltonian perturbation (\ref{pert}) has the infinitesimal form (\ref
{inf}) with 
\begin{equation}
N_{t}=L_{t}\circ B_{t}-B_{t}\circ L_{t}-L_{t}[B_{t}]\mathcal{I}.
\end{equation}
The central point of the proof that follows is the fact that the observable $
\rho ^{-1}N_{t}^{\dagger }[\rho ]$, which appears on the right hand side of (\ref{rep}),
can be expressed in terms of the 
stochastic derivative (associated
with the unperturbed process) of the observable $B_{t}$. 
\begin{eqnarray}
\rho _{t}^{-1}N_{t}^{\dagger }[\rho _{t}]&&=B_{t}\rho _{t}^{-1}L^{\dagger
}[\rho _{t}]-\rho _{t}^{-1}\circ L^{\dagger }\circ \rho
_{t}[B_{t}]-L_{t}[B_{t}]\nonumber \\
&&=-(L_{t}+L_{t}^{\ast })[B_{t}]=\left( \frac{d_{-}}{dt
}-\frac{d_{+}}{dt}\right) [B_{t}],  \label{siou}
\end{eqnarray}
where the third equality comes from (\ref{d+d-}). We can rewrite this
observable by adding a term proportional to $\left( 2\frac{d}{dt}-\frac{d_{-}
}{dt}-\frac{d_{+}}{dt}\right) B$ (which is exactly equal to zero), and then,
for all $\alpha $, we get 
\begin{equation}
\rho _{t}^{-1}N_{t}^{\dagger }[\rho _{t}]=\left( 1-\alpha \right) \frac{d_{-}
}{dt}B_{t}-(1+\alpha )\frac{d_{+}}{dt}B_{t}+2\alpha \frac{d}{dt}B_{t}.
\end{equation}
Now, by using the response relation, (\ref{rep}) and the time derivative of a
correlation function, (\ref{dercor}), we find the family, indexed by $\alpha $
, of equivalent \textbf{GFDT}. 
\begin{eqnarray}
&&\left. \frac{\delta \left\langle A_{t}(x_{t})\right\rangle ^{\prime }}{
\delta k_{s}}\right| _{k=0}\nonumber \\
&&=\left( 1-\alpha \right) \partial
_{s}\left\langle B_{s}(x_{s})A_{t}(x_{t})\right\rangle -(1+\alpha
)\left\langle \frac{d_{+}B_{s}}{ds}(x_{s})A_{t}(x_{t})\right\rangle +2\alpha
\left\langle \frac{dB_{s}}{ds}(x_{s})A_{t}(x_{t})\right\rangle. \nonumber \\
\end{eqnarray}
Two particular cases of $\alpha $ exist in the literature:

\begin{itemize}
\item  \textbf{\ }$\alpha =0$ \ : \textbf{First GFDT } 
\begin{eqnarray}
\left. \frac{\delta \left\langle A_{t}(x_{t})\right\rangle ^{\prime }}{
\delta k_{s}}\right| _{k=0} &=&\partial _{s}\left\langle
B_{s}(x_{s})A_{t}(x_{t})\right\rangle -\left\langle \frac{d_{+}B_{s}}{ds}
(x_{s})A_{t}(x_{t})\right\rangle  \\
&=&\partial _{s}\left\langle B_{s}(x_{s})A_{t}(x_{t})\right\rangle
-\left\langle \left( \partial _{s}B_{s}\right)
(x_{s})A_{t}(x_{t})\right\rangle -\left\langle \left( LB\right)
_{s}(x_{s})A_{t}(x_{t})\right\rangle. \nonumber \\  \label{TFD1}
\end{eqnarray}
\end{itemize}

In the usual case of Hamiltonian perturbation of a jump process or an
overdamped Langevin process, with (\ref{hjp}), we find $B_{t}=\frac{\beta 
}{2}O_{t}$ and then 
\begin{eqnarray}
\frac{2}{\beta }\left. \frac{\delta \left\langle A_{t}(x_{t})\right\rangle
^{\prime }}{\delta k_{s}}\right|_{k=0}&&= \partial _{s}\left\langle
O_{s}(x_{s})A_{t}(x_{t})\right\rangle -\left\langle \frac{d_{+}O_{s}}{ds}
(x_{s})A_{t}(x_{t})\right\rangle \\
&&=\partial _{s}\left\langle O_{s}(x_{s})A_{t}(x_{t})\right\rangle
-\left\langle \left( \partial _{s}O_{s}\right)
(x_{s})A_{t}(x_{t})\right\rangle -\left\langle \left( LO\right)
_{s}(x_{s})A_{t}(x_{t})\right\rangle, \nonumber \\
\end{eqnarray}
which was first written down in \cite{Cug1} for diffusion process with
additive noise and recently in \cite{Bae1,Bae2,Lip1,Liu2,Sei2} for jump
process and overdamped Langevin process. The 
equilibrium limit (\ref{ufdt}) is a bit obscure; it may be seen by noting that one has $
\left\langle \left(LO\right)(x_{s})A(x_{t})\right\rangle =\left\langle A(x_{s})\left(LO\right)(x_{t})\right\rangle =\partial_{t}\left\langle A(x_{s})O(x_{t})\right\rangle $.
However, there exists physical interpretation of the new
term $\left\langle \left( LB\right) _{s}(x_{s})A_{t}(x_{t})\right\rangle $
as the ``frenetic term'' \cite{Bae1,Bae2}.

\begin{itemize}
\item  $\alpha =-1$ : \textbf{Second} \textbf{GFDT }
\end{itemize}

\begin{equation}
\frac{1}{2}\left. \frac{\delta \left\langle A_{t}(x_{t})\right\rangle
^{\prime }}{\delta k_{s}}\right| _{k=0}=\partial _{s}\left\langle
B_{s}(x_{s})A_{t}(x_{t})\right\rangle -\left\langle \frac{dB_{s}}{ds}
(x_{s})A_{t}(x_{t})\right\rangle ,  \label{TFD2}
\end{equation}
which has the advantage that the effect of the
nonequilibrium
character of the unperturbed state is just in the second term on the right hand side.
For a diffusion process, this \textbf{GFDT} can be written \cite
{Fal2,Che3,Che4}, with (\ref{dspd}), as 
\begin{equation}
\frac{1}{2}\left. \frac{\delta \left\langle A_{t}(x_{t})\right\rangle
^{\prime }}{\delta k_{s}}\right| _{k=0}=\partial _{s}\left\langle
B_{s}(x_{s})A_{t}(x_{t})\right\rangle -\left\langle \left( \partial
_{s}B_{s}\right) (x_{s})A_{t}(x_{t})\right\rangle -\left\langle \left(
V_{s}^{c}.\nabla B_{s}\right) (x_{s})A_{t}(x_{t})\right\rangle .
\end{equation}
This {\bf GFDT} was experimentally checked in \cite{Gom1}. In the usual case of Hamiltonian perturbation of a pure jump process or a
overdamped Langevin process, with (\ref{hjp}), we find 
\begin{equation}
\frac{1}{\beta }\left. \frac{\delta \left\langle A_{t}(x_{t})\right\rangle
^{\prime }}{\delta k_{s}}\right| _{k=0}=\partial _{s}\left\langle
O_{s}(x_{s})A_{t}(x_{t})\right\rangle -\left\langle \frac{dO_{s}}{ds}
(x_{s})A_{t}(x_{t})\right\rangle .  \label{TFD2'}
\end{equation}

\subsubsection{Example of jump Langevin equation (\ref{sjlec}).}
\label{numericalFDT}

Here, we want to numerically verify the \textbf{GFDT}\ (\ref{TFD2}) for a
Markov  
process which mixes jump and diffusion. We consider the stochastic dynamics (\ref{jlec}) with $H=a\sin \theta$ and the same values for the parameters as those considered in (\ref
{sjlec}); $a=0.87s^{-1}$, $G=0.85s^{-1}$, $\beta = 0.8s$, and $r=0.001$. The process is supposed to be at time $t\leq 0$ in
the 
stationary state with $\rho (\theta )$ given in Fig. \ref{fig5}(b). Then, 
suddenly, at $t=0$, we consider a static perturbation of the Hamiltonian according to
\begin{equation}
H^{\prime }=H-k\sin \theta =\left( a-k\right) \sin \theta .
\end{equation}
We saw in the last section that this perturbation is of the form (\ref{pert}
), with 
\begin{equation}
h(\theta )=\exp \left( \frac{\beta }{2}k\sin \theta \right) .
\end{equation}
\ 

We checked numerically  
the time integrated version of the \textbf{FDT} (\ref{TFD2'}
) around steady state for $A=B=\sin \theta $. 
\begin{equation}
\frac{1}{\beta }\frac{\partial }{\partial k}\left. \left\langle \sin \theta
_{t}\right\rangle ^{\prime }\right| _{k=0}=\left\langle \sin ^{2}\theta
_{t}\right\rangle -\left\langle \sin \theta _{0}\sin \theta
_{t}\right\rangle -\int_{0}^{t}ds\left\langle C(\theta _{s})\sin \theta
_{t}\right\rangle.
\label{GFDTcheck}
\end{equation}

The form of the observable $C(\theta )\equiv \frac{d\sin \theta _{t}}{dt}
(\theta )$ is found with the help of (\ref{ddjp}) as

\begin{eqnarray}
C(\theta ) &&=\left( -a\cos (\theta )+G-\frac{1}{\beta }\left( \partial
_{\theta }\ln \rho \right) \right) \cos \theta \nonumber \\
&&+\frac{\left( \int d\theta ^{\prime }W(\theta ,\theta ^{\prime })\left(
\sin \left( \theta ^{\prime }\right) -\sin (\theta )\right) \right)}{2}\nonumber \\
&&-\frac{\rho
^{-1}(\theta )\left( \int d\theta ^{\prime }\left( \sin \left( \theta
^{\prime }\right) -\sin (\theta )\right) \rho (\theta ^{\prime })W(\theta
^{\prime },\theta )\right) }{2}, 
\end{eqnarray}
\ 

\bigskip 
\noindent with $W(\theta ,\theta ^{\prime })$ given by (\ref{sftr}) 
\begin{equation}
W(\theta ,\theta ^{\prime })=r\exp \left( -\frac{\beta a}{2}\left( \sin
(\theta ^{\prime })-\sin (\theta )\right) \right). 
\end{equation}

\begin{figure}[h!]
\begin{center}
\includegraphics[width=80mm]{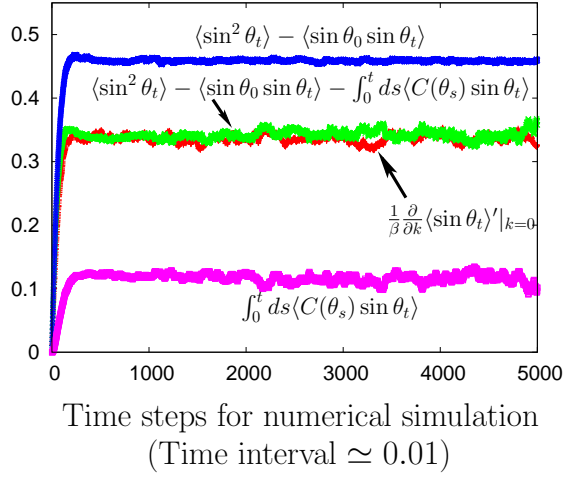}
\end{center}
\caption{Based on our results from numerical simulations of the dynamics (\ref{jlec}) with $H=a \sin \theta$, the figure shows the different terms in the integrated version of the {\bf GFDT} (\ref{GFDTcheck}). Here, $a=0.87s^{-1}$, $G=0.85s^{-1}$, $\beta = 0.8s$, and the parameter $r=0.001$. One observes a satisfactory agreement between the left hand side $\frac{1}{\beta}\frac{\partial}{\partial k}\langle \sin \theta_{t}\rangle^{\prime }| _{k=0}$ and the right side $\left\langle \sin ^{2}\theta_{t}\right\rangle -\left\langle \sin \theta _{0}\sin \theta _{t}\right\rangle -\int_{0}^{t}ds\left\langle C(\theta _{s})\sin \theta
_{t}\right\rangle$ of the {\bf GFDT} (\ref{GFDTcheck}), which therefore verifies the theorem.}
\label{fig7}
\end{figure}

Figure \ref{fig7} shows results from our numerical simulations for the various terms in the integrated version of the {\bf GFDT} (\ref{GFDTcheck}). In the figure, one can observe a satisfactory agreement between the left hand side and the right side of (\ref{GFDTcheck}), thereby verifying the {\bf GFDT}.

\subsection{Fluctuation-Dissipation Theorem for a more general class of
perturbation}

We will now consider a larger class of perturbation than (\ref{pert}) such the perturbation can be expressed in terms of two
family of non-homogeneous functions $h_t$ and $h_t^{\prime }$, as 
\begin{equation}
L_{t}^{\prime }=h_{t}^{-1}\circ L_{t}\circ h_{t}^{\prime }-h_{t}^{-1}\circ
L_{t}[h_{t}^{\prime }],  \label{genperni}
\end{equation}
which specializes to the Hamiltonian perturbation (\ref{pert}) when $
h=h^{\prime }$. We will see in the following subsections two physical
perturbations, the time change and the thermal perturbation, which belong to
this class. In the case of infinitesimal perturbation, 
\begin{equation}
h_{t}=1+k_{t}B_{t}+O(h^{2})\text{ \ \ and \ }h_{t}^{\prime
}=1+k_{t}B_{t}^{\prime }+O(h^{2}),\text{ \ }
\end{equation}
we find that this perturbation has the form (\ref{inf}) with 
\begin{equation}
N_{t}=-B_{t}\circ L_{t}+L_{t}\circ B_{t}^{\prime }-L_{t}[B_{t}^{\prime }].
\label{perinfgen}
\end{equation}
For a pure discontinuous process, the perturbation (\ref{genperni}) implies
for the transition rates, $W_{t}^{\prime }=h_{t}^{-1}\circ W_{t}\circ h_{t}^{\prime }$,
and was considered in \cite{Die1} and more recently in \cite{Mae3} by taking $h_{t}(x)=\exp (\mu \beta k_{t}O(x))$ and $h_{t}^{\prime }(x)=\exp (\gamma \beta k_{t}O(x))$,
where $O(x)$ is an observable, and $\mu $ and $\gamma $ two real numbers. As in (
\ref{siou}), we show that the observable $\rho _{t}^{-1}N_{t}^{\dagger
}[\rho _{t}]$, which appears on the right hand side of (\ref{rep}), can be expressed in
terms of the 
stochastic derivative of $B_{t}$ and $B_{t}^{\prime }$
as 
\begin{eqnarray*}
\rho ^{-1}N^{+}[\rho ] &&=-\rho ^{-1}\circ L^{\dagger }\circ \rho \lbrack
B]+B^{\prime }\rho ^{-1}L^{\dagger }[\rho ]-L[B^{\prime }]\nonumber \\
&&=-L^{\ast }\left[ B
\right] -B\rho ^{-1}L^{\dagger }[\rho ]+B^{\prime }\rho ^{-1}L^{\dagger
}[\rho ]-L[B^{\prime }] \\
&&=\left\{ 
\begin{array}{c}
\left( \partial _{t}+\rho ^{-1}L^{\dagger }[\rho ]\right) (-B+B^{\prime })+
\frac{d_{-}B}{dt}-\frac{d_{+}B^{\prime }}{dt} \\ 
\text{or},~~~\left( \partial _{t}+\rho ^{-1}L^{\dagger }[\rho ]\right)
(-B+B^{\prime })+\frac{d_{-}\left( B^{\prime }+B\right) }{dt}-2\frac{
dB^{\prime }}{dt}.
\end{array}
\right.
\end{eqnarray*}
We obtain then the generalization of the first \textbf{GFDT} (\ref{TFD1}) 
\begin{eqnarray}
&&\left. \frac{\delta \left\langle A_{t}(x_{t})\right\rangle ^{\prime }}{
\delta k_{s}}\right| _{k=0} \nonumber \\
&&=\partial _{s}\left\langle
B_{s}(x_{s})A_{t}(x_{t})\right\rangle -\left\langle \frac{d_{+}B_{s}^{\prime
}}{ds}(x_{s})A_{t}(x_{t})\right\rangle \nonumber \\
&&+\left\langle \left( \left( \partial
_{s}+\rho _{s}^{-1}L_{s}^{\dagger }[\rho _{s}]\right) (B_{s}^{\prime
}-B_{s})\right) (x_{s})A_{t}(x_{t})\right\rangle,
\label{TFDpergen}
\end{eqnarray}
and the second \textbf{GFDT} (\ref{TFD2}) \ 
\begin{eqnarray}
&&\left. \frac{\delta \left\langle A_{t}(x_{t})\right\rangle ^{\prime }}{
\delta k_{s}}\right| _{k=0} \nonumber \\
&&=\partial _{s}\left\langle \left( B_{s}^{\prime
}+B_{s}\right) (x_{s})A_{t}(x_{t})\right\rangle -2\left\langle \frac{
dB_{s}^{\prime }}{ds}(x_{s})A_{t}(x_{t})\right\rangle\nonumber \\
&&+\left\langle \left(
\left( \partial _{s}+\rho _{s}^{-1}L_{s}^{\dagger }[\rho _{s}]\right)
(B_{s}^{\prime }-B_{s})\right) (x_{s})A_{t}(x_{t})\right\rangle.
\label{TFDpergen'}
\end{eqnarray}
We see that in the Hamiltonian perturbation class (i.e., $B=B^{\prime }$), we
recover the \textbf{GFDT} (\ref{TFD1},\ref{TFD2}).\ 

\subsubsection{Around Steady State}

We will now restrict to the case where the observable does not have explicit
time dependence (i.e., $A_{t}=A$, $B_{t}=B,B_{t}^{\prime }=B^{\prime })$,
and the unperturbed state is a steady state. Then the first \textbf{GFDT} (
\ref{TFDpergen}) becomes \ 
\begin{equation}
\left. \frac{\delta \left\langle A(x_{t})\right\rangle ^{\prime }}{\delta
k_{s}}\right| _{k=0}=\partial _{s}\left\langle B(x_{s})A(x_{t})\right\rangle
-\left\langle \frac{d_{+}B^{\prime }}{ds}(x_{s})A(x_{t})\right\rangle ,
\label{TFDpergeni}
\end{equation}
and the second (\ref{TFDpergen'}) becomes 
\begin{equation}
\left. \frac{\delta \left\langle A(x_{t})\right\rangle ^{\prime }}{\delta
k_{s}}\right| _{k=0}=\partial _{s}\left\langle \left( B+B^{\prime }\right)
(x_{s})A(x_{t})\right\rangle -2\left\langle \frac{dB^{\prime }}{ds}
(x_{s})A(x_{t})\right\rangle .  \label{TFDpergeni'}
\end{equation}
In the case where the steady state is of 
equilibrium-type (i.e. $\frac{d}{ds}=\partial _{s}$), this
last relation (\ref{TFDpergeni'}) simplifies to
the  
form 
\begin{equation}
\left. \frac{\delta \left\langle A(x_{t})\right\rangle ^{\prime }}{\delta
k_{s}}\right| _{k=0}=\partial _{s}\left\langle \left( B+B^{\prime }\right)
(x_{s})A(x_{t})\right\rangle .  \label{geneq}
\end{equation}

\subsubsection{Time change for a homogeneous Markov process \protect\cite
{Dem1}}

\label{timec} An example of perturbation which belongs to the generalized
class (\ref{genperni}) but not to the Hamiltonian perturbation class (\ref
{pert}) is when we consider the change of clock as follows. 
\begin{equation}
t^{f}(s)\equiv \int_{0}^{s}du\exp \left( f_{u}\left( x_{u}\right) \right) ,
\label{tc}
\end{equation}
where $f_{u}$ is an observable. It is proved in \cite{Dem1} (proposition
3.1) that the process $x_{s}^{\prime }=x_{t^{f}(s)}$ is still Markov  
with
a generator $L_{t}^{^{\prime }}=\exp (-f_{t})L$. In the case of infinitesimal perturbation $f_{u}(x)=k_{u}B(x)$, $L_{t}^{^{\prime }}=L-k_{t}BL$,
which is of the form (\ref{perinfgen}) with $B^{\prime }=0$ so that the FDT (
\ref{TFDpergen}, \ref{TFDpergen'}) takes the form 
\begin{equation}
\left. \frac{\delta \left\langle A(x_{t})\right\rangle ^{\prime }}{\delta
k_{s}}\right| _{k=0}=\partial _{s}\left\langle B(x_{s})A(x_{t})\right\rangle
-\left\langle \left( \rho _{s}^{-1}L^{\dagger }[\rho _{s}]B\right)
(x_{s})A(x_{t})\right\rangle .  \label{fdttc}
\end{equation}
In the case of an unperturbed system in the steady state, (\ref
{TFDpergeni},\ref{TFDpergeni'}) become the usual \textbf{FDT}. 
\begin{equation}
\left. \frac{\delta \left\langle A(x_{t})\right\rangle ^{\prime }}{\delta
k_{s}}\right| _{k=0}=\partial _{s}\left\langle B(x_{s})A(x_{t})\right\rangle
,  \label{fdttcs}
\end{equation}
which is a result of \cite{Dem1}. We want to emphasize that, for this type
of perturbation, we obtain the usual \textbf{FDT} (without correction)\ for
an unperturbed state which is a general  
nonequilibrium steady
state.

\subsubsection{Thermal Perturbation Pulse : Change of temperature for
equilibrium overdamped Langevin process}

\label{thermalp} A famous example in physics for a perturbation which is not
of Hamiltonian type is thermal perturbation. Let us consider a system whose
dynamics is governed by (\ref{sys}), with $G=0$ and homogeneous Hamiltonian,
and the perturbed system which results from the change of the bath
temperature $\,\beta _{t}^{-1}=(1+k_{t})\beta ^{-1}.$ We can easily prove
that 
\begin{equation}
N=\frac{1}{\beta }\nabla _{i}\circ \Gamma ^{ij}\circ \nabla _{j},
\end{equation}
which is of the form of the general infinitesimal perturbation (\ref
{perinfgen}) with $B^{\prime }=\frac{\beta H}{2}$ \ and $B=\frac{\beta H}{2}
-1.$ This can be easily seen by using the formula (\ref{carc}) for the
''carre du champs'': 
\begin{equation}
L\circ B^{\prime }-L[B^{\prime }]=B^{\prime }L+\Gamma (B^{\prime },.)=B^{\prime }L+\frac{2\Gamma ^{ij}}{\beta }
\left( \nabla _{i}B^{\prime }\right) \nabla _{j}=\frac{\beta H}{2}L-L+N.
\end{equation}
The formula (\ref{geneq}) then takes the form $\left. \frac{\delta \left\langle A_{t}(x_{t})\right\rangle ^{\prime }}{
\delta k_{s}}\right| _{k=0}=\partial _{s}\left\langle \left( \beta
H-1\right) (x_{s})A_{t}(x_{t})\right\rangle$,
which implies the equilibrium form 
\begin{equation}
\frac{1}{\beta }\left. \frac{\delta \left\langle A_{t}(x_{t})\right\rangle
^{\prime }}{\delta k_{s}}\right| _{h=0}=\partial _{s}\left\langle
H(x_{s})A_{t}(x_{t})\right\rangle .
\end{equation}
In particular, we obtain the usual \textbf{FDT} for the energy \cite{Risk}. 
\begin{equation}
\frac{1}{\beta }\left. \frac{\delta \left\langle H(x_{t})\right\rangle
^{\prime }}{\delta k_{s}}\right| _{k=0}=\partial _{s}\left\langle
H(x_{s})H(x_{t})\right\rangle .  \label{fdttp}
\end{equation}

\section{Two families of non-perturbative 
extensions of the Fluctuation-Dissipation Theorem}

It is well understood since the discovery of the \textbf{FR} that they may
be viewed as extensions to the non-perturbative regime of the Green-Kubo and
Onsager relations which are usually valid within the linear response
description in the vicinity of 
equilibrium \cite{Gall2,LebowSp}. A detailed proof was given in \cite{Che1} 
that the
Jarzynski equality gives
the usual\textbf{\ FDT} when expanded to second order in the response
field. In \cite{Fal2}, it was proved that this correspondence is still true
around an unperturbed state which is  
stationary  
but 
out of
equilibrium)
. This is proved by doing a Taylor expansion of a special 
Crooks theorem to first order in the response field. Finally, in 
\cite{Che4}, general \textbf{FR} were exhibited which are global versions of
the \textbf{GFDT} for 
nonequilibrium diffusion, or, of the \textbf{
FDT} for energy resulting from a thermal perturbation. 
We  
introduce in Section (\ref{expmaa}) a first family of 
exponential martingales
which is a natural object associated with the perturbation (\ref{pert}), and
show in Section (\ref{fdtteme}) that these are global version of general 
\textbf{GFDT} (\ref{TFD1},\ref{TFD2}). Section (\ref
{fludoob}) presents the martingale property of functionals 
which appear  
in the fluctuation relations and it shows  
their relation  
to the 
exponential martingales introduced in Section (\ref{expmaa}).
Along the way, we  
prove  
the \textbf{FR} along the lines of the proof given
below  
for the 
exponential martingale by  
a comparison to the backward process generated by the
Doob $h$-transform of the adjoint generator $L^{\dagger }$.  

\subsection{New family of exponential 
martingales 
naturally related to \textbf{GFDT} }

\label{expmaa}

\subsubsection{Introduction}
 We come back to the perturbation (\ref{pert}) of the generator, 
\begin{equation}
L_{t}^{h}\equiv h_{t}^{-1}\circ L_{t}\circ h_{t}-h_{t}^{-1}L_{t}\left[ h_{t}
\right] ,  \label{pert'}
\end{equation}
but this time we will not restrict to the regime where $h_{t}$ is infinitesimal. We prove in Appendix (\ref{annpt}) that the Markov process associated with the generators $L_{t}^{h}$ and $L_{t}$ are related through the  
functional $\exp (-\mathbf{Z}_{s}^{h,t})$ by  
\begin{equation}
P_{s}^{h,t}(x,y)=\mathbf{E}_{s,x}\left[ \delta (x_{t}-y)\exp (-\mathbf{Z}_{s}^{h,t})\right],  \label{lienpt}
\end{equation}
where 
\begin{equation}
\exp (-\mathbf{Z}_{s}^{h,t}\left[ x\right] )=h_{s}^{-1}(x_{s})\exp \left(
-\int_{s}^{t}du\left( h_{u}^{-1}L_{u}\left[ h_{u}\right] +h_{u}^{-1}\partial
_{u}h_{u}\right) (x_{u})\right) h_{t}(x_{t}).  \label{eZ}
\end{equation}
The functional $\exp (-\mathbf{Z}_{s}^{h,t})$ is
multiplicative:
\begin{equation}
\exp (-\mathbf{Z}_{s}^{h,t})=\exp (-\mathbf{Z}_{s}^{h,u})\,
\exp (-\mathbf{Z}_{u}^{h,t})
\end{equation}
for $s\leq u\leq t$.  
The perturbation (\ref{pert'}) is a particular case 
of the transformation of a Markov process by multiplicative 
functionals \cite{Blu1,Ito}.
It is a generalization of the Doob $h$-transform, which is 
\begin{equation}
P_{s}^{h,t}(x,y)=h_{s}^{-1}(x)P_{s}^{t}(x,y)h_{t}(y),
\end{equation}
in the case where $h_{t}(x)$ is the  
space-time harmonic function, i.e. $\frac{d_+h}{dt}=0$.

Thanks to the Markovian structure 
of the trajectory measure, 
the relation (\ref{lienpt}) for the transition probability is  
equivalent (as proved in Appendix \ref{annme}) to the relation between 
the expectations of functionals of the 
paths from time $s$ to time $t$ for the perturbed and 
the unperturbed processes, 
\begin{equation}
\mathbf{E}_{s,x}^{h}\left[ F_{\left[ s,t\right] }\right] =\mathbf{E}_{s,x}\left[ F_{\left[ s,t\right] }\exp (-\mathbf{Z}_{s}^{h,t}[x])\right] ,  \label{martex}
\end{equation}
where $\mathbf{E}_{s,x}^{h}\left[ {~~}\right] $ denotes expectation for 
the process with  
generators $L_{t}^{h}.$ 

Finally, we can also formulate (\ref{martex}) by requiring that the 
perturbed process 
with generators $L_{t}^{h}$ 
and
trajectory measure $M_{\mu_{0,}\left[ s,t\right] }^{h}$\footnote{Note 
that the measure $\mu_{0}(dx)=\rho_0(x)dx$ is the measure
at initial time $t=0$ and 
not at time $s$.},  can be obtained from the unperturbed process with 
trajectory measure $M_{\mu_{0,}\left[ s,t\right] }$ by using the likelihood 
ratio process (the Radon-Nikodym density): 
\begin{equation}
M_{\mu_{0,}\left[ s,t\right] }^{h}[x]=M_{\mu_{0,}\left[ s,t\right]
}[x]{\frac{\rho^{h} _{s}}{\rho _{s}}(x_{s})} \exp (-\mathbf{Z}_{s}^{h,t}[x]),  \label{martexp}
\end{equation}
with $\rho_{s}(x)=\int dy \rho_{0}(y)P_{0}^{s}(y,x)$  the instantaneous 
density of the original process and $\rho^{h}_{s}(x)=\int dy 
\rho_{0}(y)P_{0}^{h,s}(y,x)$ the instantaneous density of the 
$h$-transformed process.

We could not find the general result (\ref{eZ},\ref{martex},\ref{martexp}) in the mathematics literature, but many very closely related results do exist. The subcase of (\ref{martex},\ref{martexp}) where $h_{t}$ is time-homogeneous (i.e., $\partial _{t}h_{t}=0)$) was treated long time ago by Kunita in \cite{Kun1} and was revisited recently in the articles \cite{Pal1} and \cite{Dia1}. In our context of the {\bf FDT}, the extension to $\partial _{t}h_{t}\neq 0$ is 
essential. But more than generalizing to 
$\partial _{t}h_{t}\neq 0,$ the main interest in the Appendices (\ref{annpt},\ref{annme})\ is to prove (\ref{martex}) from theoretical physics perspective. We recall also that for a diffusion process, the perturbed generator (\ref{pert'}) is obtained by adding the term $d_{t}\nabla \left( \ln h_{t}(x) \right) $ to the drift (see (\ref{adddrift})). Then, the proofs in Appendices (\ref{annpt},\ref{annme}) are also a theoretical physicist's proofs of the Girsanov theorem 
for a diffusion process \cite{Rev1} (for this type 
of change of drift).
 
\subsubsection{Martingale properties of the functional $\exp (-\mathbf{Z}_{s}^{t}$)   }
The multiplicative functional $\exp (-\mathbf{Z}_{s}^{h,t}\left[ x\right] )$ is an \textbf{exponential martingale} with respect to 
the 
natural $\sigma$-algebra filtration $\mathcal{F}_{t}=\sigma(x_s,s\leq t)$  
representing the increasing flow of information. This fact can be seen 
by first noting that (\ref{martex}) (with $F=1$) implies
the normalization condition 
\begin{equation}
\mathbf{E}_{s,x}\left[ \exp \left(-\mathbf{Z}_{s}^{h,t}\left[ x\right] \right)\right] =1,
\label{ajr}
\end{equation}
which, thanks to the multiplicative structure of 
$\exp (-\mathbf{Z}_{s}^{h,t}\left[ x\right] )$, 
yields 
\begin{equation}
\mathbf{E}\left[ \exp (-\mathbf{Z}_{s}^{h,t}\left[ x\right] )|\mathcal{F}_{u}\right] =\exp (-\mathbf{Z}_{s}^{h,u}\left[ x\right] )\mathbf{E}_{x_{u},u}\left[ \exp (-\mathbf{Z}_{u}^{h,t}\left[ x\right] )\right] =\exp (-\mathbf{Z}_{s}^{h,u}\left[ x\right] ),
\label{promar}
\end{equation}
for $s\leq u\leq t.$

\subsubsection{Fluctuation-Dissipation Theorem as Taylor expansion of the exponential martingale identity (\ref{martex})}

\label{fdtteme} In the infinitesimal case, $h_{t}(x)=1+k_{t}B_{t}(x)+O(k^{2})$,
Taylor expansion of the subcase of (\ref{martex}) with one-point functional $
F_{\left[ s,t\right] }[x]=A_{t}(x_{t})$, namely, 
\begin{equation}
\left\langle A_{t}(x_{t})\right\rangle ^{h}=\left\langle \left( h_{s}\right)
^{-1}(x_{s})\exp \left( \int_{s}^{t}du\left( -h_{u}^{-1}L_{u}\left[ h_{u}
\right] -h_{u}^{-1}\partial _{u}h_{u}\right) (x_{u})\right)
h_{t}(x_{t})A_{t}(x_{t})\right\rangle,
\end{equation}
gives, to the first order in $\left[ k\right] $, 
\begin{eqnarray}
& & \left\langle A_{t}(x_{t})\right\rangle ^{h}-\left\langle A_{t}(x_{t})\right\rangle -\left\langle \left( h_{s}\right) ^{-1}(x_{s})A_{t}(x_{t})\right\rangle      -\left\langle \left(h_{t}\right) (x_{t})A_{t}(x_{t})\right\rangle \nonumber \\
&&+\int_{s}^{t}du\left\langle \frac{
d_{+}h_{u}}{du}(x_{u})A_{t}(x_{t})\right\rangle +O(k^{2})=0.
\label{dell}
\end{eqnarray}

To find the first {\bf GFDT} (\ref{TFD1}) from (\ref{dell}), we use the direct
differentiation formula $\frac{d_{+}h_{u}}{du}=\frac{d_{+}\left(
k_{u}B_{u}\right) }{du}=\left( \partial _{u}k_{u}\right) B_{u}+k_{u}\frac{
d_{+}\left( B_{u}\right) }{du}$ obtaining the relation

\begin{equation}
\left. \frac{\delta \left\langle A_{t}(x_{t})\right\rangle ^{\prime }}{
\delta k_{u}}\right| _{k=0}-\partial _{u}\left\langle
B_{u}(x_{u})A_{t}(x_{t})\right\rangle +\left\langle \frac{d_{+}B_{u}}{du}
(x_{u})A_{t}(x_{t})\right\rangle =0,
\end{equation}
which is the  
{\bf GFDT} (\ref{TFD1}) (note the equivalence 
of the two notations $
\left\langle {\ }\right\rangle ^{\prime }$ and $\left\langle {\ }\right\rangle
^{h})$). To find the second \textbf{GFDT} (\ref{TFD2}) from (\ref{dell}), we use
the formula $d_{+}=2d-d_{-}$, which gives 
\begin{eqnarray}
& &\left\langle A_{t}(x_{t})\right\rangle ^{h}-\left\langle A_{t}(x_{t})\right\rangle -\left\langle \left( h_{s}\right) ^{-1}(x_{s})A_{t}(x_{t})\right\rangle      -\left\langle \left(h_{t}\right) (x_{t})A_{t}(x_{t})\right\rangle  \nonumber \\
& &+2\int_{s}^{t}du\left\langle 
\frac{dh_{u}}{du}(x_{u})A_{t}(x_{t})\right\rangle -\int_{s}^{t}du\left\langle \frac{d_{-}h_{u}}{du}(x_{u})A_{t}(x_{t})\right
\rangle +O(k^{2})=0.\qquad
\end{eqnarray}
Then, by using (\ref{dercor}) for the time derivative of a correlation 
function, we 
obtain 
\begin{eqnarray}
& &\left\langle A_{t}(x_{t})\right\rangle ^{h}-\left\langle A_{t}(x_{t})\right\rangle -\left\langle \left( h_{s}\right) ^{-1}(x_{s})A_{t}(x_{t})\right\rangle      -\left\langle \left(h_{t}\right) (x_{t})A_{t}(x_{t})\right\rangle+  \nonumber \\
&& 2\int_{s}^{t}du\left\langle 
\frac{dh_{u}}{du}(x_{u})A_{t}(x_{t})\right\rangle -\int_{s}^{t}du\partial
_{u}\left\langle h_{u}(x_{u})A_{t}(x_{t})\right\rangle +O(k^{2})=0.\qquad
\end{eqnarray}
Next, we use the differentiation formula $\frac{dh_{u}}{du}=\frac{d\left(
k_{u}B_{u}\right) }{du}=\left( \partial _{u}k_{u}\right) B_{u}+k_{u}\frac{
d\left( B_{u}\right) }{du}$ to get 
\begin{equation}
\left. \frac{\delta \left\langle A_{t}(x_{t})\right\rangle ^{\prime }}{
\delta k_{u}}\right| _{k=0}-2\partial _{u}\left\langle
B_{u}(x_{u})A_{t}(x_{t})\right\rangle +2\left\langle \frac{dB_{u}}{du}
(x_{u})A_{t}(x_{t})\right\rangle =0,
\end{equation}
which is (\ref{TFD2}).  
This 
gives a second independent proof of (\ref{TFD1},
\ref{TFD2}). 
It also shows that 
the above 
exponential martingales
are natural global versions of the 
 \textbf{GFDT}. We will 
discuss in the next section another  
well known global version, namely, the 
Fluctuation Relations (\textbf{FR}).

\subsection{
Family of exponential martingales 
related to \textbf{GFDT} through Fluctuation 
Relations}
\label{fludoob} 
\subsubsection{Introduction to 
Fluctuation Relations}

Roughly speaking,  
\textbf{FR} 
may be obtained by comparing
the expectation of functionals of 
trajectories of the system  
and of 
reversed trajectories of the  so-called backward
system\footnote{It is important to underline that this backward process 
is not unique. We can also call it a comparison process
} (denoted by an index $r$)
. More precisely, let us denote by $M^{r}_{\mu_{0}^{r},\left[0,T\right]}$ the trajectorial measure of the backward process which is initially distributed with the measure $\mu_{0}^{r}(dx)\equiv \rho_{0}^{r}(x)dx$.  Next, we define the path-wise time inversion $R$ at fixed time $T$\footnote{
 This 
was also studied in the probabilistic literature, but 
with time  
$T$ that 
could be  
random  
\cite{Chu1}. },
which acts on the space of trajectories according to $R\left[x\right]_{t}=\left[x\right]_{T-t}$\footnote{For simplicity, we neglect the case where the time inversion acts non-trivially on the space by an involution. Such a situation arises for Hamiltonian systems (see \cite{Che1}) where the  
involution is
\,$
(q,p)\rightarrow(q,-p) 
.$}. 
This allows us to  introduce the (push-forward) measure  $R_{\ast}\left(M_{\mu_{0},\left[0,T\right]}\right)$, which is the measure of the trajectory but traversed in the backward sense.  
We  then introduce the action functional $\mathbf{W_{0}^{T}}$  
through the Radon Nykodym derivative
of the image measure of the backward system $R_{\ast}\left(M^{r}_{\mu_{0}^{r},\left[0,T\right]}\right)$
with respect to the trajectorial measure  $M_{\mu_{0},\left[0,T\right]}$ of the forward system (initially distributed with the measure   $\mu_{0}(dx)\equiv \rho_{0}(x)dx$):\footnote{
We assume that the measures 
$M_{\mu_{0},\left[0,T\right]}$ and $R_{\ast}\left(M^{r}_{\mu_{0}^{r},\left[0,T\right]}\right)$ are mutually absolutely continuous.}

\begin{equation}
R_{\ast}\left(M^{r}_{\mu_{0}^{r},\left[0,T\right]}\right) \equiv \exp\left(-\mathbf{W}_{0}^{T}\right)M_{\mu_{0},\left[0,T\right]}.
 \label{actionfunct}
\end{equation}

Equivalently, we can write this relation in the form of the
Crooks  
theorem  
\cite{Maes,Crooks2,LebowSp,Sch1,Che1,Liu1,Sei2} 
asserting that for all trajectory 
functionals $F_{\left[ 0,T\right] }$,
\begin{equation}
\left\langle F_{\left[ 0,T\right] }\circ R\right\rangle ^{r}
=\left\langle F_{\left[ 0,T\right] } 
\exp \left(-\mathbf{W}_{0}^{T}\right)\right\rangle.
\label{fluC}
\end{equation}
Finally, by the substitution $F_{\left[ 0,T\right]}\left[x\right] \rightarrow F_{\left[ 0,T\right]}\left[x\right]\delta(x_{0}-y)]\delta(x_{T}-x),$ we find the equivalent relation, 
\begin{eqnarray}
&&\rho_{0}^{r}(x)\mathbf{E}_{0,x}^{r}\left[F_{\left[ 0,T\right] }\circ R\left[x\right]\delta(x_{T}-y)\right]\\
&&\hspace{1.5cm}=\rho_{0}(y)\mathbf{E}_{0,y}\left[F_{\left[ 0,T\right] }\left[x\right]\exp\left(-\mathbf{W}_{0}^{T}\left[x\right]\right)\delta(x_{T}-x)\right]. 
\label{entpro} 
\end{eqnarray}

\noindent Due to the freedom in choosing the initial 
forward measure $\mu _{0}$ or
the backward measure $\mu _{0}^{r}$, it is possible to  
identify
the action
functional  
with various thermodynamic quantities 
like the work performed on the
system or the fluctuating entropy creation $\mathbf{\sigma}_{0}^{T}\left[x\right]$ with respect to the inversion $r$ . This latter quantity is obtained when $\mu_{0}^{r}(dx)=\rho_{0}^{r}(x)dx=\rho _{T}(x)dx \equiv \int dy \rho_{0}(y)P_{0}^{T}(y,x)dx $.
We can also obtain the functional entropy production in the environment, $\mathbf{J}_{0}^{T}$, by choosing $\mu_{0}(dx)=\mu_{0}^{r}(dx)=dx$, because then the difference from the entropy creation is the boundary term 
$\ln \left(\rho_{0}(x_{0})\right)-\ln \left( \rho_{t}(x_{t})\right) $,
which gives the change in the instantaneous entropy of the
process.

Let us observe the similarity between 
(\ref{martex},\ref{martexp}
) and (\ref{entpro},\ref{actionfunct}); $\exp (-\mathbf{Z}_{s}^{h,t}\left[ x
\right] )$ and $\exp (-\mathbf{
W}^{T}_{0}\left[ x\right] )$ are exponential functionals of
the Markov  
process.  
However $\exp (-\mathbf{W}^{T}_{0}\left[ x\right] )$
is not a forward martingale in the generic case because (\ref{entpro}) implies that 
\begin{equation}
\text{\ }\mathbf{E}_{0,y}\left[ \exp (-\mathbf{W}^{T}_{0})\right] =\frac{\int dx \rho
_{0}^{r}(x)P_{0}^{r,T}(x,y)}{\rho _{0}(y)}.
\end{equation}
Moreover, 
\begin{equation}
 \exp (-\mathbf{W}^{0}_{0})(y) =\frac{ \rho_{0}^{r}(y)}{\rho _{0}(y)},
 \label{W0}
 \end{equation}
  and then $\mathbf{E}_{0,y}\left[ \exp (-\mathbf{W}^{T}_{0})\right]\neq \exp (-\mathbf{W}^{0}_{0})(y) $,
except  
in the case where  $\rho _{0}^{r}$ is an invariant 
density of the backward dynamics.
The fact that $\exp (-\mathbf{W}^{T}_{0}\left[ x\right] )$ 
is not a forward
martingale does not prevent us from obtaining the 
Jarzynski equality 
\cite{Crooks2,Jarz}, 
\begin{equation}
\left\langle \exp (-\mathbf{W}^{T}_{0})\right\rangle =1,  \label{JE}
\end{equation}
which is a direct subcase of (\ref{fluC}). We will show in the next section 
that there is nevertheless a martingale interpretation of the action 
functional  
and the Jarzynski equality is one of
its consequences.    

The Jarzynski relation (\ref{JE}) implies two important results. First, the Jenssen inequality allows to obtain the Second 
Law of Thermodynamics, 
\begin{equation}
\left\langle \mathbf{W}_{0}^{T}\right\rangle \geq0.
\end{equation}
Second, the Markov inequality  \cite{Dur} gives an upper bound on the 
probability of ``transient deviations'' 
from the Second 
Law: 
\begin{equation}
\mathbf{P}\left(\exp\left(-\mathbf{W}_{0}^{T}\right)\geq \exp(L) \right)\leq\frac{\left\langle \exp\left(-\mathbf{W}_{0}^{T}\right)\right\rangle }{\exp(L)} \text{\,\,\,then\,\,\,} \mathbf{P}\left(\mathbf{W}_{0}^{T}\leq-L\right)\leq\exp\left(-L\right).  \label{travio}
\end{equation}

\subsubsection{Martingale properties of the action functional }

We 
noted in the last section that the  
functional $\exp(-\mathbf{W}_{0}^{T})$
is not a martingale with respect to the time $T$ of inversion. In order 
to  
unravel  
its links  with the martingale 
theory, we  
shall define 
a functional similar to $\mathbf{W}_{0}^{T}$,  but with a lower time 
indices different from $0$ and a upper time indices different from $T$. 
This will be done through the comparison of the  
trajectorial measure $M_{\mu_{0},\left[s,t\right]}$   
for the forward system 
on the subinterval $\left[s,t\right]$ of $\left[0,T\right]$ and the push 
forward by the time inversion 
$R_{\ast}\left(M^{r}_{\mu_{0}^{r},\left[T-t,T-s\right]}\right)$ 
of the trajectorial measure  
for the backward system on the sub interval 
$\left[T-t,T-s\right]$\footnote{Then the two measures deal with the ``same part'' of the trajectory. } 
\begin{equation}
R_{\ast}\left(M_{\mu_{0}^{r},\left[T-t,T-s\right]}^{r}\right)=\exp\left(-\mathbf{W}_{s}^{t}\right)M_{\mu_{0},\left[s,t\right]}.
\label{afunst}
\end{equation}

Proceeding as in the last section (\ref{fluC}), we can write the 
Crooks-type  
theorem for all functional $F_{\left[s,t\right]}$ of the trajectories from $s$ to $t$,
\begin{equation}
\left\langle F_{\left[s,t\right]}\circ R\right\rangle ^{r}=\left\langle F_{\left[s,t\right]}exp\left(-\mathbf{W}_{s}^{t}\right)\right\rangle, 
\label{funst}
\end{equation}
or, equivalently,  
\begin{eqnarray}
&&\rho_{T-t}^{r}(x)\mathbf{E}_{T-t,x}^{r}\left[F_{\left[s,t\right]}\circ R\left[x\right]\delta(x_{T-s}-y)\right]\cr
&&\hspace{1.5cm}=\rho_{s}(y)\mathbf{E}_{s,y}\left[F_{\left[s,t\right]}\left[x\right]exp\left(-\mathbf{W}_{s}^{t}\left[x\right]\right)\delta(x_{t}-x)\right],
\label{funstt}
\end{eqnarray}
with \,\,\,$\rho_{s}^{r}(x)=\int dy \rho_{0}^{r}(y)P_{0}^{r,s}(y,x)$\,\,\, and \,\,\, $\rho_{s}(x)=\int dy \rho_{0}(y)P_{0}^{s}(y,x).$
Finally, this includes also  a Jarzynski type relation 
(\ref{JE})\footnote{Note that  $
 \exp (-\mathbf{W}^{s}_{s})(y) =\frac{ \rho_{T-s}^{r}(y)}{\rho _{s}(y)}$ 
and this  
seems to 
contradict (\ref{W0}) in the limit 
$s\rightarrow 0$. The resolution of the paradox is that 
expression here is obtained
by the limit  
at fixed $T$: $\lim\limits_{s\rightarrow 0}
\lim\limits_{t\rightarrow s}\mathbf{W}^{t}_{s}$, while (\ref{W0}) results 
from a different limiting procedure: $\lim\limits_{T\rightarrow 0} 
\lim\limits_{s\rightarrow 0} \mathbf{W}^{T}_{s}$.}:  
\begin{equation}
\left\langle \exp (-\mathbf{W}^{t}_{s})\right\rangle =1.
\label{JE'} 
\end{equation}

For studying the martingale properties of $\exp (-\mathbf{W}^{t}_{s})$, it is important to note that this functional is not strictly multiplicative.     
For $0\leq s \leq u\leq t \leq T $, the Markov properties\footnote{ 
 \begin{equation}
 M_{\mu_{0},\left[s,t\right]}\left[dx\right]=\frac{M_{\mu_{0},\left[s,u\right]}\left[dx\right]M_{\mu_{0},\left[u,t\right]}\left[dx\right]}{\rho_{u}(x_{u})dx_{u}}, \nonumber
\end{equation}
 and  
\begin{equation}
R_{\ast}\left(M_{\mu_{0}^{r},\left[T-t,T-s\right]}^{r}\right)\left[dx\right]={\frac{R_{\ast}\left(M_{\mu_{0}^{r},\left[T-t,T-u\right]}^{r}\right)\left[dx\right]R_{\ast}\left(M_{\mu_{0}^{r},\left[T-u,T-s\right]}^{r}\right)\left[dx\right]}{\rho^{r}_{T-u}(x_{u})dx_{u}}}, \nonumber
\end{equation}
where the right hand sides describe the disintegration
of the left-hand-side measures with respect to the map evaluating
trajectories at time $u$.}
imply  the ``multiplicative'' law for the action functional: 
\begin{equation}
\exp\left(-\mathbf{W}_{s}^{t}\left[dx\right]\right)=\exp\left(-\mathbf{W}_{s}^{u}\left[dx\right]\right)\exp\left(-\mathbf{W}_{u}^{t}\left[dx\right]\right)\frac{\rho_{u}}{\rho_{T-u}^{r}}(x_{u}).
\label{mul} 
\end{equation}  
This allows to 
introduce two functionals, 
\begin{equation}
\mathbf{A}_{s}^{t}\left[dx\right] \equiv \frac{\rho_{s}}{\rho_{T-s}^{r}}(x_{s}) \exp\left(-\mathbf{W}_{s}^{t}\left[dx\right]\right) \text{\;\; and  \;\;} \mathbf{R}_{s}^{t}\left[dx\right] \equiv  \exp\left(-\mathbf{W}_{s}^{t}\left[dx\right]\right) \frac{\rho_{t}}{\rho_{T-t}^{r}}(x_{t})
\label{defAR}
\end{equation}
with the strict multiplicative property: 
\begin{equation}
\mathbf{A}_{s}^{t}=\mathbf{A}_{s}^{u}\mathbf{A}_{u}^{t} \text{ \;\; and  \;\;} \mathbf{R}_{s}^{t}=\mathbf{R}_{s}^{u}\mathbf{R}_{u}^{t}.
\label{mul'}  
\end{equation} 
For these two functionals, the relation (\ref{funstt}) implies that
\begin{equation}
\mathbf{E}_{s,y}\left[\mathbf{A}_{s}^{t}\left[x\right]\delta(x_{t}-x)\right]= \frac{\rho^{r}_{T-t}(x)}{\rho_{T-s}^{r}(y)}\mathbf{E}_{T-t,x}^{r}\left[\delta(x_{T-s}-y)\right],
\end{equation}
  and
\begin{equation}
\mathbf{E}_{s,y}\left[\mathbf{R}_{s}^{t}\left[x\right]\delta(x_{t}-x)\right]= \frac{\rho_{t}(x)}{\rho_{s}(y)}\mathbf{E}_{T-t,x}^{r}\left[\delta(x_{T-s}-y)\right],
\end{equation}
yielding the Jarzynski-type 
relations, 
\begin{equation} 
 \mathbf{E}_{s,y}\left[\mathbf{A}_{s}^{t}\right]= 1,
 \label{jart}   
\end{equation} 
and 
\begin{equation} 
 \left\langle \mathbf{R}_{s}^{t}\left[x\right]\delta(x_{t}-x)\right\rangle=\rho_{t}(x), \text{ \;\;i.e. \;\; } \mathbf{E}\left(\left.\mathbf{R}_{s}^{t}\right|x_{t}=x\right)=1.
 \label{jars} 
\end{equation}  
Then, by using the multiplicative property (\ref{mul'}) and the 
relation (\ref{jart}), we see that $\mathbf{R}_{s}^{t}$ is a forward 
martingale with respect to the 
natural filtration   $\mathcal{F}_{t},$ :  
\begin{equation}
\mathbf{E}\left[  \mathbf{A}_{s}^{t}|\mathcal{F}_{u}\right] =\mathbf{A}_{s}^{u},
\end{equation}
for $s\leq u\leq t$. Similarly, by using the multiplicative property (\ref{mul'}) and the 
relation (\ref{jars}), we see that $\mathbf{R}_{s}^{t}$ is a backward 
martingale\footnote{A backward martingale \cite{Del,Dur} 
is dual to forward martingale,  
in the sense that its
expectation in the past, given the knowledge accumulated  
in the future, is its present value.} with respect to 
the filtration $\mathcal{G}_{s}=\sigma(x_{v},v\geq s)$ which describes 
the future of the process, 
  
\begin{equation}
\mathbf{E}\left[  \mathbf{R}_{s}^{t}|\mathcal{G}_{u}\right] =\mathbf{R}_{u}^{t},
\end{equation}
for $s\leq u\leq t$.
From the definition (\ref{defAR}), we deduce that the action functional  $\exp\left(-\mathbf{W}_{s}^{t}\right)$ with  $0\leq s \leq u\leq t < T $  is a forward martingale with respect to upper indices $t$\footnote{This is a little 
tricky because we proved in the last section  
that $\exp\left(-\mathbf{W}_{s}^{T}\right)$  is not a 
forward martingale  
with respect to $T$. What happens for $t=T$ is that  
a change of $T$ implies also a change of the time 
inversion  
$R$ which  
breaks the martingale property.} 
\textbf{and} a backward martingale with respect to lowest indices $s$.

This gives a martingale interpretation 
of the Jarzynski equality (\ref{JE},\ref{JE'}).    
One possible application is to improving the upper bound 
of the probability of ``transient deviations'' of the 
Second 
Law.  The Doob inequality \cite{Del,Dur} for forward 
martingales gives a stronger upper bound than the  
Markov inequality (\ref{travio}),
\begin{eqnarray}
\mathbf{P}\left(\sup_{t \text{\;with \;}0\leq s\leq t\leq T}\left(\exp\left(-\mathbf{W}_{s}^{t}\right)\right)\geq \exp(L)\right)\leq\frac{\left\langle \exp\left(-\mathbf{W}_{s}^{t}\right)\right\rangle }{\exp(L)},
\end{eqnarray}
and then we obtain the relation
\begin{eqnarray}
\mathbf{P}\left(\inf_{t \text{\;with \;}0\leq s\leq t\leq T}\mathbf{W}_{s}^{t}\leq-L\right)\leq\exp\left(-L\right)&&.
\label{dev2}
\end{eqnarray}
\subsubsection{
Action functional $\mathbf{W}_{s}^{t}$  
and the time reversed process}
\label{ref}
It is proved in the probability literature \cite{Chu1,Dyn2,Fol1,Hau1,Nel,Nel2} 
that the time-reversed process, $RX
_{t}
\equiv X_{t^{\ast }}$ (with $t^{\ast }=T-t$),  
is also a Markov process. 
By using the results of Section  \ref{Cot}, and more specifically, the expression of the cogenerator,  (\ref{Lb}), we can deduce that the Markov  
generator of  
the time-reversed process is   
\begin{equation}
L_{t}^{CR}=L_{t^{\ast }}^{\ast } \equiv \rho
_{t^{\ast }}^{-1}\circ L_{t^{\ast }}^{\dagger }\circ \rho _{t^{\ast }}-\rho _{t^{\ast }}^{-1}L_{t^{\ast }}^{\dagger
}[\rho _{t^{\ast }}]\mathcal{I}.
\label{totinv}
\end{equation}
Choosing this process as the backward process was called complete reversal in \cite{Che1}, and this explains the index ``CR'' on the left hand side. 
We remark that the instantaneous density of the time-reversed system 
is related to that of the original system by   
$\rho _{t}^{CR}\equiv \rho _{t^{\ast }}$. 
Denoting by $M_{\mu_{0},\left[s,t\right]}^{r \circ CR}$ the trajectorial  
measure of the time reversal of the backward process initially 
distributed with 
the measure $\mu^{r}_{0}$, we have the tautological formula: 
\begin{equation}
R_{\ast}\left(M_{\mu_{0}^{r},\left[T-t,T-s\right]}^{r}\right)=M_{\mu_{T}^{r},\left[s,t\right]}^{r \circ CR}.
\end{equation}
This allows to obtain an expression 
for the action 
functional $\mathbf{W}_s^t$ from (\ref{afunst}) 
without push-forward $R_{\ast}$
:
\begin{equation}
M_{\mu_{T}^{r},\left[s,t\right]}^{r \circ CR} =\exp\left(-\mathbf{W}_{s}^{t}\right)M_{\mu_{0},\left[s,t\right]}.
\label{afunst'}
\end{equation}
This expression will be used in the next section, but it also 
allows an easy proof of the assertion that 
$\exp\left(-\mathbf{W}_{s}^{t}\right)$ is a forward martingale in 
$t$ and a backward martingale in $s$.

\subsubsection{Class of action functional $\mathbf{W}_{s}^{t}$  (\ref{afunst})  which are in relation with the exponential martingale  $\mathbf{Z}_{s}^{t}$ (\ref{eZ}) }

We consider here the case where
the backward process is given by the generalized 
Doob $f$-transform of
the adjoint generator $L^{\dagger }$ composed with an inversion of the time:

\begin{equation}
L_{u^{\ast }}^{r,f}=f_{u}^{-1}\circ L_{u}^{\dagger }\circ
f_{u}-f_{u}^{-1}L_{u}^{\dagger }\left[ f_{u}\right]; ~~~~u^{\ast }=T-u.  \label{bacgen}
\end{equation}

\noindent We shall denote by $\mathbf{W}_{s}^{f,t}$
the action functional associated with this choice of the backward process. 

Using the definition of the total inversion  (\ref{totinv}) and after some algebra, one may show that

\begin{equation}
\left(L_{u}^{r,f}\right)^{CR }=\left(\rho_{u^{\ast }}^{r}\right)^{-1}f_{u} \circ L_{u} \circ \rho_{u^{\ast }}^{r}f_{u}^{-1}-\left(\rho_{u^{\ast }}^{r}\right)^{-1}f_{u}  L_{u}[\rho_{u^{\ast }}^{r}f_{u}^{-1}]\mathcal{I}.
\end{equation}
So,  
for $h_{u}\equiv \rho_{u^{\ast }}^{r}f_{u}^{-1}$, we have
\begin{equation}
\left(L_{u}^{r,f}\right)^{CR }=L_{u}^{h},
\end{equation}
upon using the definition of the generalized Doob $h$-transform  (\ref{pert}). 

Finally, by comparing the relations (\ref{afunst'}) and  (\ref{martexp}),  we find the link between the two families of functionals\footnote{Note that $ \rho^{h} _{s}=\rho^{r} _{s ^{\ast}}$}:  
\begin{equation}
\exp\left(-\mathbf{W}_{s}^{f,t}\right)={\frac{\rho^{r} _{s^{\ast}}}{\rho _{s}}(x_{s})}\exp\left(-\mathbf{Z}_{s}^{h,t}\right),
\end{equation}
with $h_{u}\equiv \rho_{u^{\ast }}^{r}f_{u}^{-1}$. 
Moreover, this relation allows to obtain from  (\ref{eZ})  an explicit expression for $\exp\left(-\mathbf{W}_{s}^{f,t}\right)$:  
\begin{eqnarray}
 \exp\left(-\mathbf{W}_{s}^{f,t}\right)&=&{\frac{\rho^{r} _{s^{\ast}}}{\rho _{s}h_{s}}(x_{s})}\exp \left(-\int_{s}^{t}du\left( h_{u}^{-1}L_{u}\left[ h_{u}\right] +h_{u}^{-1}\partial
_{u}h_{u}\right) (x_{u})\right) h_{t}(x_{t})\qquad
 \nonumber \\
&= & \frac{f_{s}}{\rho _{s}}(x_{s})\exp \left(-\int_{s}^{t}du\left( h_{u}^{-1}L_{u}\left[ h_{u}\right] +h_{u}^{-1}\partial
_{u}h_{u}\right) (x_{u})\right) \frac{\rho^{r} _{t^{\ast}}}{f_{t}}(x_{t}) \nonumber \\
&= & \frac{f_{s}}{\rho _{s}}(x_{s})
\exp \left( -\int_{s}^{t}du\left(
f_{u}^{-1}L_{u}^{\dagger }\left[ f_{u}\right] -f_{u}^{-1}
\left( \partial_{u}f_{u}\right) \right)\right) 
\frac{\rho^{r} _{t^{\ast}}}{f_{t}}(x_{t}),\quad
\label{fin}
\end{eqnarray}
with, \,as before, \,\,\,$\rho_{s}(x)=\int dy \rho_{0}(y)P_{0}^{s}(y,x)$\,\,\, and \,\,\, $\rho^{r}_{t^*}(x)
=\int dy \rho^{r}_{0}(y)P_{0}^{r,t^*}(y,x)$\footnote{
The last equality in  (\ref{fin}) results from  the following algebra: 
\begin{eqnarray}
h_{u}^{-1}\partial_{u}h_{u}&=&
\left(\rho_{u^{\ast }}^{r}\right)^{-1}f_{u}  \partial_{u} \left(  
\rho_{u^{\ast }}^rf_{u}^{-1}\right) \nonumber \\
&=&
-f_{u}^{-1}  \partial_{u}  f_{u}+\left(\rho_{u^{\ast }}^{r}\right)^{-1}  
\partial_{u} \left(  \rho_{u^{\ast }}^r\right)
\nonumber \\ 
&=&-f_{u}^{-1}  \partial_{u}  f_{u}-\left(\rho_{u^{\ast }}^{r}
\right)^{-1}\left(L_{u^{\ast }}^{r,f}\right)^{\dagger }[ 
\rho_{u^{\ast }}^{r}]\nonumber\\
&=& - f_{u}^{-1}  \partial_{u}  f_{u}+f_{u}^{-1}L_{u}^{\dagger }\left[ f_{u}\right]-\left(\rho_{u^{\ast }}^{r}\right)^{-1}f_{u}  L_{u}[\rho_{u^{\ast }}^{r}f_{u}^{-1}].\nonumber
\end{eqnarray}
}.
In particular, the action functional with $s=0$ and $t=T$,  which results 
for the choice $\rho _{0}=f_{0}$ and $
\rho _{0}^{r }=f_{T}$, is then

\begin{equation}
\mathbf{W}^{f,T}_{0}=\int_{0}^{T}du\left( f_{u}^{-1}L_{u}^{\dagger }\left[ f_{u}
\right] -f_{u}^{-1}\left( \partial _{u}f_{u}\right) \right), \label{W}
\end{equation}

The form (\ref{bacgen}) taken for
the backward generator may be justified by showing 
that it allows  
to recover the
forms of time inversion usually taken in the probability or physics
literature.

\begin{itemize}
\item  First, we remark that the usual Doob $f$-transform 
corresponds to the case where
we take for $f_{t}$ the PDF (\ref{pdf}) of the forward process (i.e., $
\partial _{t}f_{t}-L_{t}^{\dagger }.
f_{t}=0)$. Then, we recognize
using
formula (\ref{Lb}) that $L_{t^{\ast }}^{r}=L_{t}^{\ast }$ and then this 
backward process is the one obtained from the
original one by the ``complete reversal'' 
considered in Section (\ref{ref}). This implies
with (\ref{sgb-Lb}) that  
\begin{eqnarray}
&&P_{s}^{ \ast t }=
\overrightarrow{\exp }\left( \int_{s}^{t}du\left(L_{u}^{\ast }\right) ^{\dagger }\right)=
\overrightarrow{\exp }\left( \int_{s}^{t}du\left(L^{CR}_{u^{\ast }}\right) ^{\dagger }\right)\nonumber\\
&&\hspace{2cm}=
\overrightarrow{\exp }\left( \int_{t^{\ast }}^{s^{\ast }}du\left(L^{CR}_{u}\right) \right)^{\dagger }=
\left( P_{t^{\ast }}^{s^{\ast },CR}\right) ^{\dagger
}.
\end{eqnarray}

Finally, with (\ref{2pb}), we obtain the generalized detailed balance, 
\begin{equation}
\rho _{s}(x)P_{s}^{t}(x,y)=\rho _{t}(y)P_{t^{\ast }}^{s^{\ast },CR}(y,x).
\end{equation}
One may show that $\rho _{t}^{CR}\equiv \rho _{t^{\ast }}$ is the
instantaneous density of the backward process and that the corresponding
current operator (\ref{JV}) satisfies the relation 
\begin{equation}
J_{t}^{CR}\,=\,-\,J_{t^{\ast }}\,,  \label{jjr}
\end{equation}
which is very satisfying physically.  
This choice, however, corresponds to the vanishing of the
functional $\mathbf{W}^{t}_{s}$  
and of the entropy creation (equal to it
due to the choice $\rho _{0}^{CR}=\rho _{T})$. 

\item  Another useful choice of time inversion, called the current reversal
in \ \cite{CHCHJAR,Che1}, is based on the choice $f_{t}=\pi _{t}\,$, where $
\,\pi _{t}$ is the $\,$accompanying density (\ref{AD}). One can show that $\pi _{t}^{r}\equiv \pi _{t^{\ast }}$ is then the accompanying
density for the backward process. If we associate with 
the 
accompanying density the current 
operator, by analogy with (\ref{JV}), 
\begin{equation}
J_{t}\equiv \pi _{t}\circ L_{t}-L_{t}^{\dagger }\circ \pi _{t},
\end{equation}
we can easily show that  
still $J^r_t=-J_{t^\ast}$. The functional (\ref{fin})  now takes the form 
\begin{equation}
\exp (-\mathbf{W}_{s}^{f=\pi,t})=\frac{\pi_{s}}{\rho _{s}}(x_{s})\exp \left( \int_{s}^{t}du\left( \partial
_{u}\ln \pi_{u}\right)(x_{u}) \right)  \left(\frac{\pi_{t}}{\rho^{r}_{T-t}}\right)^{-1}(x_{t}).  \label{Wex}
\end{equation}
Moreover, the choice of initial density  $\rho_{0}=\pi_0$ and 
and  $\rho^{r}_{0}=\pi^r_{0}
=\pi_{T}$, implies 
that
\begin{equation}
\mathbf{W}_0^{f=\pi,T}\,=\,-\,\int\limits_{0}^{T}(\partial
_{u}\ln \pi _{u})(x_{u})\,du\ \equiv\,
\mathbf{W}_0^{T,ex},  \label{WTC}
\end{equation}
where the index ``ex'' stands for ``excess'' \cite{Che1,Oon1,Sei}.
\thinspace The 
Jarzynski equality 
(\ref{JE}) for this case was
first proved for a one-dimensional diffusion process in \cite{HatSas} and
then for Markov chains \cite{Che5,Ge}, general 
diffusion processes \cite{Che1,Liu1},
and pure jump processes \cite{Liu3}. We see here 
that these \textbf{FR} are
true for general Markov processes, including stochastic equation with
Poisson noise (\ref{GPN}) or with Levy noise. This is an optimistic result
for the generality of \textbf{FR} in the context of the proof in \cite
{Tou1,Bau1} that the 
Gallavotti-Cohen relation for the work is
broken for a particle in a harmonic potential  
subject to a Poisson or
Levy noise. Moreover, in the case of the jump Langevin equation (\ref{lepe}
), we have the normalized accompanying density $\pi _{t}=\exp (-\beta \left( H_{t}-F_{t}\right) )$ (where $F_t$ is the 
free energy)  and then 
\begin{equation}
\mathbf{W}_{0}^{T,ex}\,=\,\beta
\,\int\limits_{0}^{T}(\partial _{u}
H_{u})(x_{u})\,du-\beta \left(
F_{T}-F_{0}\right) .
\end{equation}
So, in this case, 
the finite time {\bf FR} (\ref{fluC})
for the dissipative work performed 
on the system
is 
valid.

\item   
For 
diffusion processes, it was shown in \cite{Che1} that
to obtain a sufficiently flexible notion of time inversion, we 
should allow
for a non-trivial behavior of the modified drift $\widehat{u}_{t}$ 
(see \cite{Che1}) under the
time-inversion by dividing it into two parts: 
\begin{equation}
\widehat{u}_{t}=\widehat{u}_{t,+}+u_{t,-}.
\end{equation}
Here $\widehat{u}_{t,+}$ transforms as a vector field under time inversion,
i.e., $\widehat{u}_{t^{\ast },+}^{r}=+\,\widehat{u}_{t,+}$, while 
$u_{t,-}$ transforms as a pseudo-vector field, i.e., $u_{t^{\ast
},-}^{r}=-\,u_{t,-}$. The random field $\eta_{t}$ may be transformed
with either of the two rules: $\eta _{t^{\ast }}^{r}=\pm \eta
_{t} $. It can be shown \cite{Che4} that the choice of the vector field
part which allows us to obtain the backward generator given by (\ref{bacgen}
) is 
\begin{equation}
\widehat{u}_{t,+}=\frac{d_{t}}{2}\nabla \left( \ln  f_{t}
\right). 
\end{equation}
This is the choice  
made to obtain formula $(22)$ in \cite{Che4} 
in order to find \textbf{
FR} that are global versions of \textbf{GFDT} in the context of a Langevin
process, and we find (\ref{W}) as formula 
$(24)$ in \cite{Che4}.
\end{itemize}

\subsubsection{Fluctuation-Dissipation Theorem as Taylor expansion of fluctuation relation for the class of functional $\exp (-\mathbf{W}_{s}^{f,t})$  }

For completeness, we recall here the proof, done in 
\cite{Che4} for a diffusion process, that the family of {\bf FR} (\ref{fluC}) 
with
(\ref{W}) are also global versions of the \textbf{GFDT} (\ref{TFD1},\ref
{TFD2}). More precisely, they are global versions of the fundamental relation of the linear response theory (\ref{rep}
) which, as explained in Section (\ref{fdth}), implies the {\bf GFDT} (\ref{TFD1},
\ref{TFD2}).

For the dynamics of the perturbed systems (\ref{pert}), we consider the
fluctuation relation (\ref{fluC}) with the functional (\ref{W}) written for $
f_{t}$ (with $f_{t}(x)=1+k_{t}A_{t}(x)+O(k^{2})$) as the mean instantaneous density $\rho _{t}$ of the unperturbed
system (with $f=1 \equiv k=0$). The functional (\ref{W}) becomes 
\begin{equation}
\mathbf{W}^{f,T}_{0}=\int_{0}^{T}k_{s}\left( \rho _{s}{}^{-1}N_{s}^{\dagger }\left[\rho
_{s}]\right]\right) (x_{s})\,ds,
\end{equation}
where $N$ is defined in (\ref{inf}) . Let us now write a particular case of (
\ref{fluC}) for a single time functional $F[x]=A^{a}(x_{t})\equiv A_{t}^{a}$
($\,0<t<T$): 
\begin{equation}
\big\langle A_{t}\,\mathrm{\exp }\left( -\mathbf{W}^{f,T}_{0}\right) \big\rangle
^{f}\,=\,\big\langle A_{T-t}\big\rangle^{f,r}\hspace{0.025cm}.\text{ \ }
\label{Crtop}
\end{equation}
The first order Taylor expansion, 
\begin{equation}
\exp (-\mathbf{W}^{f,T}_{0})=1+\int_{0}^{T}k_{s}\,\left( (\rho
_{s}{}^{-1}N_{s}^{\dagger }[\rho _{s}]\right) (x_{s})\,ds\,+\,\mathcal{O}
(k^{2}),
\end{equation}
in (\ref{Crtop}) gives the relation 
\begin{eqnarray}
\big\langle A_{t}\,\big\rangle+\int k_{s}\frac{\delta }{\delta k_{s}}\Big|
_{k=0}\,\left\langle A_{t}\right\rangle ^{f
}\,ds\,-\int_{0}^{T}k_{s}\left\langle \rho _{s}{}^{-1}N^{\dagger }[\rho
_{s}](x_{s})A_{t}(x_{t})\right\rangle ds\,+\,\mathcal{A}(k^{2})\nonumber\\
=\,
\big\langle A_{T-t}\big\rangle^{f,r}.\qquad
\end{eqnarray}
Due to the form of the considered inversion (\ref{bacgen}), the right hand side has a functional dependence only on $\{k_{T-u},T-u<T-t\}$,  
i.e. on $\{k_{u},u>t\}.$ So,
if we apply $\,\frac{\delta }{\delta k_{s}}|_{k=0}\,$ for $\,0<s\leq t\,$ to
the last identity, we obtain the relation (\ref{rep}).

\section{Conclusions}

We have shown that the kinematics of a Markov process, namely, the 
local velocity (\ref{vf},\ref{vb},\ref{sov}) and the 
derivatives (\ref{d}), allow to develop a unified approach to obtain recent 
\textbf{GFDT} in the context of fairly general Markovian 
evolutions (Section (
\ref{fdth})). We have also  
elucidated the form of the usual 
perturbation (
\ref{pert}) used for \textbf{FDT} by showing its similarity 
 to the Doob $h$-transform well known
in the probabilistic literature. We
also presented examples where the physical perturbation is more general,
e.g. given by a time change (\ref{tc}) or 
by a thermal perturbation (Section (\ref{thermalp}
)). We derived the \textbf{GFDT} for these examples (\ref{fdttc},\ref{fdttcs}
,\ref{fdttp}). In this paper, we have also presented a class of 
the  
exponential  martingale functionals 
(\ref{eZ}), which represents
an alternative to {\bf FR} as a non-perturbative 
extension of \textbf{GFDT} (Section \ref{fdtteme}). 
Moreover, we
established in Section 6.2.4 a direct  
link between this family of functionals and the {\bf FR}.    
We showed that the {\bf FR} also involve a family
of martingales which for a fairly general class of
{\bf FR}, including several classes discussed in the literature, 
coincides with exponential martingales. This class of {\bf FR}
was obtained 
by comparison of the original Markov process to 
the backward process whose generators (\ref{bacgen}) are generalized 
Doob transforms
for the adjoints of the original generators.  
In the process, we improved 
the classical upper bound for ``transient deviations''  
from the Second  
Law (\ref{dev2}). Our hope is that, despite
lack of rigor from the mathematical perspective, this article will serve as
a bridge between 
nonequilibrium physics and probability theory.

\begin{acknowledgements}
The authors thank Andre Barato, Michel Bauer, Gregory
Falkovich,Krzysztof Gawedzki, Ori Hirschberg, Kirone Mallick and David Mukamel for discussions and comments on the manuscript. Special thanks are due to Krzysztof 
Gawedzki for his valuable input on Section 6.2. RC acknowledges
support of the Koshland Center for Basic Research. SG thanks Freddy Bouchet, Thierry Dauxois, David Mukamel and Stefano Ruffo for encouragement. SG also acknowledges the Israel Science Foundation (ISF) for supporting his research at the Weizmann Institute and the contract ANR-10-CEXC-010-01, Chaire d'Excellence for supporting his research at Ecole Normale Sup\'{e}rieure, Lyon.
\end{acknowledgements}

\appendix

\section{Proof of the relation (\ref{dercor})}

\label{dcor} By taking the derivative of the relation (\ref{coro}) with
respect to $t~(>s)$, we get 
\begin{eqnarray*}
\partial _{t}\left\langle U_{s}\left( x_{s}\right) V_{t}\left( x_{t}\right)
\right\rangle &=&\int dxdyU_{s}(x)\rho _{s}(x)\left( \partial
_{t}P_{s}^{t}(x,y)\right) V_{t}(y)+\int dxdyU_{s}(x)\rho
_{s}(x)P_{s}^{t}(x,y)\left( \partial _{t}V_{t}(y)\right) \\
&=&\left\langle U_{s}\left( x_{s}\right) \left( \partial
_{t}V_{t}+L_{t}(V_{t})\right) \left( x_{t}\right) \right\rangle
=\left\langle U_{s}\left( x_{s}\right) \frac{d_{+}V_{t}}{dt}\left(
x_{t}\right) \right\rangle .
\end{eqnarray*}
In deriving the second line, we use the forward Kolmogorov equation (\ref{FK}
). Now, by taking the derivative of (\ref{coro}) with respect to $s~(<t)$,
and by using the definition of the cogenerator (\ref{sgb-Lb}), we get 
\begin{eqnarray*}
&&\partial _{s}\left\langle U_{s}\left( x_{s}\right) V_{t}\left( x_{t}\right)
\right\rangle \\
&=&\int dxdyU_{s}(x)\left( \partial _{s}P_{s}^{\ast
t}(x,y)\right) \rho _{t}(y)V_{t}(y)+\int dxdy\left( \partial
_{s}U_{s}(x)\right) P_{s}^{\ast t}(x,y)\rho _{t}(y)V_{t}(y) \\
&=&-\int dxdyU_{s}(x)\left( \left( L_{s}^{\ast }\right) ^{\dagger
}P_{s}^{\ast t}\right) (x,y)\rho _{t}(y)V_{t}(y)+\int dxdy\left( \partial
_{s}U_{s}(x)\right) P_{s}^{\ast t}(x,y)\rho _{t}(y)V_{t}(y) \\
&=&-\int dxdydzU_{s}(x)L_{s}^{\ast }(z,x)P_{s}^{\ast t}(z,y)\rho
_{t}(y)V_{t}(y)+\int dxdy\left( \partial _{s}U_{s}(x)\right) P_{s}^{\ast
t}(x,y)\rho _{t}(y)V_{t}(y) \\
&=&-\int dydz\left( L_{s}^{\ast }U_{s}\right) (z)P_{s}^{\ast t}(z,y)\rho
_{t}(y)V_{t}(y)+\int dxdy\left( \partial _{s}U(s,x)\right) P_{s}^{\ast
t}(x,y)\rho _{t}(y)V_{t}(y) \\
&=&\left\langle \left( \partial _{s}U_{s}-L_{s}^{\ast }U_{s}\right) \left(
x_{s}\right) V_{t}\left( x_{t}\right) \right\rangle =\left\langle \frac{
d_{-}U_{s}}{ds}\left( x_{s}\right) V_{t}\left( x_{t}\right) \right\rangle .
\end{eqnarray*}

\section{Proof of the relation (\ref{Lstardiff})}

\label{cogpd} The formal adjoint of the generator (\ref{gend}) of a
diffusion process is given by 
\begin{equation*}
L_{t}^{\dagger }=-\nabla _{i}\circ \widehat{u}_{t}^{i}+\frac{1}{2}\nabla
_{i}\circ d_{t}^{ij}\circ \nabla _{j}.
\end{equation*}
So, for all functions $f_{t}$ in\ $\mathcal{E}$, we can express the operator $
f_{t}^{-1}\circ L_{t}^{\dagger }\circ f_{t}$ as 
\begin{eqnarray*}
&&f_{t}^{-1}\circ L_{t}^{\dagger }\circ f_{t} \\
&=&f_{t}^{-1}\circ \left( -\nabla _{i}\circ \widehat{u}_{t}^{i}+\frac{1}{2}
\nabla _{i}\circ d_{t}^{ij}\circ \nabla _{j}\right) \circ
f_{t}\\
&=&f_{t}^{-1}\circ \left( -\left( \nabla _{i}\left( \widehat{u}
_{t}^{i}f_{t}\right) \right) -\widehat{u}_{t}^{i}f_{t}\circ \nabla _{i}+
\frac{1}{2}\nabla _{i}\circ \left( d_{t}^{ij}\left( \nabla _{j}f_{t}\right)
+f_{t}d_{t}^{ij}\circ \nabla _{j}\right) \right)  \notag \\
&=&f_{t}^{-1}\left( -\left( \nabla _{i}\left( \widehat{u}_{t}^{i}f_{t}
\right) \right) -\widehat{u}_{t}^{i}f_{t}\circ \nabla _{i}+\frac{1}{2}\nabla
_{i}(d_{t}^{ij}\nabla _{j}f_{t})+\frac{d_{t}^{ij}}{2}\left( \nabla
_{j}f_{t}\right) \circ \nabla _{i}+\frac{d_{t}^{ij}}{2}\left( \nabla
_{i}f_{t}\right) \circ \nabla _{j}+\frac{f_{t}}{2}\nabla _{i}\circ
d_{t}^{ij}\circ \nabla _{j}\right)  \notag \\
&=&f_{t}^{-1}L_{t}^{\dagger }[f_{t}]+L_{t}-2\left( \widehat{u}_{t}^{i}-\frac{
d_{t}^{ij}}{2}\left( \nabla _{j}\ln f_{t}\right) \right) \nabla _{i}.
\label{a1}
\end{eqnarray*}
Moreover, with $f_{t}=\rho _{t}$ (PDF (\ref{pdf})) and with (\ref{eqm}), \
we obtain 
\begin{equation*}
\rho _{t}^{-1}\circ L_{t}^{\dagger }\circ \rho _{t}-\rho _{t}^{-1}\left(
\partial _{t}\rho _{t}\right) =L_{t}-2\left( \widehat{u}_{t}^{i}-\frac{
d_{t}^{ij}}{2}\left( \nabla _{j}\ln \rho _{t}\right) \right) \nabla _{i}.
\end{equation*}
By using the definition of the cogenerator (\ref{Lb}) and of the \textbf{
hydrodynamic velocity} (\ref{jv}), we obtain the formula (\ref{Lstardiff}).

\section{Proof of the relation (\ref{rep})}

\label{linres} We start with the first-order Dyson expansion \cite{Joa} of the ordered
exponential (\ref{gen}) :\ 
\begin{equation*}
\overrightarrow{\exp }\left( \int_{0}^{t}du\left( L_{u}+k_{u}N_{u}\right)
\right) =\overrightarrow{\exp }\left( \int_{0}^{t}duL_{u}\right)
+\int_{0}^{t}ds\overrightarrow{\exp }\left( \int_{0}^{s}duL_{u}\right)
k_{s}N_{s}\overrightarrow{\exp }\left( \int_{s}^{t}dsL_{s}\right) +O(k^{2}).
\end{equation*}
Then, for the one point functional, one has 
\begin{equation*}
\left\langle A_{t}(x_{t})\right\rangle ^{\prime }=\int dxdy\rho
_{0}(x)P_{0}^{t}(x,y)A_{t}(y)+\int_{0}^{t}dsk_{s}\int dxdydzdz^{\prime }\rho
_{0}(x)P_{0}^{s}(x,y)N_{s}(y,z)P_{s}^{t}(z,z^{\prime })A_{t}(z^{\prime
})+O(k^{2}).
\end{equation*}
The response function is then given by 
\begin{eqnarray*}
\left. \frac{\delta \left\langle A_{t}(x_{t})\right\rangle ^{\prime }}{
\delta k_{s}}\right| _{k=0} &=&\int dxdydzdz^{\prime }\rho
_{0}(x)P_{0}^{s}(x,y)N_{s}(y,z)P_{s}^{t}(z,z^{\prime })A_{t}(z^{\prime }) \\
&=&\int dydzdz^{\prime }\rho _{s}(y)N_{s}(y,z)P_{s}^{t}(z,z^{\prime
})A_{t}(z^{\prime }) \\
&=&\int dzdz^{\prime }\left( N_{s}^{\dagger }\rho _{s}\right)
(z)P_{s}^{t}(z,z^{\prime })A_{t}(z^{\prime }).
\end{eqnarray*}
which is (\ref{rep}).

\section{Proof of the relation (\ref{carc})}

\label{pcarc} With the formula (\ref{gend}) for the generator of a diffusion
process, and for two arbitrary functions $f$ and $g$ on $\mathcal{E}$, one
has 
\begin{eqnarray*}
L_{t}[fg] &=&\widehat{u}_{t}^{i}\nabla _{i}[fg]+\frac{1}{2}\left( \nabla
_{i}\circ d_{t}^{ij}\circ \nabla _{j}\right) [fg]=f\widehat{u}_{t}^{i}\left(
\nabla _{i}g\right) +\widehat{u}_{t}^{i}\left( \nabla _{i}f\right) g+\frac{1
}{2}\nabla _{i}\left( d_{t}^{ij}\left( \nabla _{j}f\right)
g+d_{t}^{ij}f\left( \nabla _{j}g\right) \right) \\
&=&f\widehat{u}_{t}^{i}\left( \nabla _{i}g\right) +\widehat{u}_{t}^{i}\left(
\nabla _{i}f\right) g+\frac{1}{2}\left( \nabla _{i}\left( d_{t}^{ij}\nabla
_{j}f\right) \right) g+d_{t}^{ij}\left( \nabla _{j}f\right) \left( \nabla
_{i}g\right) +\frac{f}{2}\left( \nabla _{i}\left( d_{t}^{ij}\nabla
_{j}g\right) \right) \\
&=&fL_{t}(g)+L_{t}(f)g+d_{t}^{ij}\left( \nabla _{i}f\right) \left( \nabla
_{j}g\right) .
\end{eqnarray*}
We then obtain the formula (\ref{carc}) for the operator ''carre du
champs'': 
\begin{equation*}
\Gamma _{t}(f,g)=d_{t}^{ij}\left( \nabla _{i}f\right) \left( \nabla
_{j}g\right) .
\end{equation*}

\section{Proof of the relation (\ref{lienpt})}

\label{annpt} We start by proving the operatorial relation 
\begin{equation*}
P_{s}^{h,t}=h_{s}^{-1}\overrightarrow{\exp }\left( \int_{s}^{t}du\left(
L_{u}-h_{u}^{-1}L_{u}\left[ h_{u}\right] -h_{u}^{-1}\partial
_{u}h_{u}\right) \right) h_{t}.  \label{PPh}
\end{equation*}
First, it is easy to see that the above relation is true when $t=s$ (also,
then both the left hand side and the right hand side equal the identity).
Moreover, we now show that the two sides of the relation verify the same
differential equation. For example, the right hand side satisfies 
\begin{eqnarray*}
&&\partial _{t}\left( h_{s}^{-1}\overrightarrow{\exp }\left(
\int_{s}^{t}du\left( L_{u}-h_{u}^{-1}L_{u}\left[ h_{u}\right]
-h_{u}^{-1}\partial _{u}h_{u}\right) \right) h_{t}\right) \\
&=&h_{s}^{-1}\overrightarrow{\exp }\left( \int_{s}^{t}du\left(
L_{u}-h_{u}^{-1}L_{u}\left[ h_{u}\right] -h_{u}^{-1}\partial
_{u}h_{u}\right) \right) \circ \left( L_{t}\circ h_{t}-L_{t}\left(
h_{t}\right) -\partial _{t}h_{t}+\partial _{t}h_{t}\right) \\
&=&\left( h_{s}^{-1}\overrightarrow{\exp }\left( \int_{s}^{t}du\left(
L_{u}-h_{u}^{-1}L_{u}\left[ h_{u}\right] -h_{u}^{-1}\partial
_{u}h_{u}\right) \right) h_{t}\right) \circ \left( h_{t}^{-1}\circ
L_{t}\circ h_{t}-h_{t}^{-1}L_{t}(h_{t})\right) \\
&=&\left( h_{s}^{-1}\overrightarrow{\exp }\left( \int_{s}^{t}du\left(
L_{u}-h_{u}^{-1}L_{u}\left[ h_{u}\right] -h_{u}^{-1}\partial
_{u}h_{u}\right) \right) h_{t}\right) \circ L_{t}^{h}.
\end{eqnarray*}
It is easy to see by using the forward Kolmogorov equation that the right
hand side verifies the same equation. Now, we can apply the Feynman-Kac
formula \cite{Rev1,Str} to the right hand side of (\ref{PPh}), and we obtain
the relation (\ref{lienpt}), namely, 
\begin{equation*}
P_{s}^{h,t}(x,y)=\mathbf{E}_{s,x}\left( \delta (x_{t}-y)\exp (-\mathbf{Z}_{s}^{h,t}
\left[ x\right] )\right) ,
\end{equation*}
with the functional $\exp (-\mathbf{Z}_{s}^{t}\left[ x\right] )$ given by
the relation (\ref{eZ}).

\section{Proof of the relation (\ref{martex})}

\label{annme} We now want to prove the relation (\ref{martex}) with $F$ an
arbitrarily functional of the trajectories on $\left[ s,t\right] .$ It
suffices to check this identity for the so-called cylindrical functional: 
\begin{equation*}
F_{\left[ s,t\right] }[x]=F(x_{s},x_{t_{1}},x_{t_{2}},...,x_{t_{n}},x_{t})\text{ for }
s\leq t_{1}\leq t_{2}\leq ...\leq t_{n}\leq t.  \label{cf}
\end{equation*}
We will use the Markov structure of the trajectory measure. 
\begin{eqnarray*}
&&\mathbf{E}_{s,x}^{h}\left[ F(x_{s},x_{t_{1}},x_{t_{2}},...,x_{t_{n}},x_{t})\right] \\
&=&\int
dx_{1}dx_{2}...dx_{n}dyF(x,x_{1},x_{2},...,x_{n},y)P_{s}^{h,t_{1}}(x,dx_{1})P_{t_{1}}^{h,t_{2}}(x_{1},dx_{2})...P_{t_{n}}^{h,t}(x_{n},y)
\\
&=&\int dx_{1}dx_{2}...dx_{n}dyF(x,x_{1},x_{2},...,x_{n},y)h_{s}^{-1}(x) \\
&&\times \mathbf{E}_{s,x}\left[ \delta (x_{t_{1}}-x_{1})\exp \left(
-\int_{s}^{t_{1}}du\left( h_{u}^{-1}L_{u}\left( h_{u}\right)
+h_{u}^{-1}\partial _{u}h_{u}\right) \right) \right]
h_{t_{1}}(x_{1})h_{t_{1}}^{-1}(x_{1}) \\
&&\times \mathbf{E}_{t_{1},x_{1}}\left[ \delta (x_{t_{1}}-x_{2})\exp \left(
-\int_{t_{1}}^{t_{2}}du\left( h_{u}^{-1}L_{u}\left( h_{u}\right)
+h_{u}^{-1}\partial _{u}h_{u}\right) \right) \right] ... \\
&&\times \mathbf{E}_{t_{n},x_{n}}\left[ \delta (x_{t}-y)\exp \left(
-\int_{t_{n}}^{t}du\left( h_{u}^{-1}L_{u}\left( h_{u}\right)
+h_{u}^{-1}\partial _{u}h_{u}\right) \right) \right] h_{t}(y) \\
&=&\int dx_{1}dx_{2}...dx_{n}dyF(x,x_{1},x_{2},...,x_{n},y)h_{s}^{-1}(x)E_{s,x} 
\left[ \delta (x_{t_{1}}-x_{1})\exp \left( -\int_{s}^{t_{1}}du\left(
h_{u}^{-1}L_{u}\left( h_{u}\right) +h_{u}^{-1}\partial _{u}h_{u}\right)
\right) \right] \\
&&\times \mathbf{E}_{t_{1},x_{1}}\left[ \delta (x_{t_{1}}-x_{2})\exp \left(
-\int_{t_{1}}^{t_{2}}du\left( h_{u}^{-1}L_{u}\left( h_{u}\right)
+h_{u}^{-1}\partial _{u}h_{u}\right) \right) \right] ... \\
&&\times \mathbf{E}_{t_{n},x_{n}}\left[ \delta (x_{t}-y)\exp \left(
-\int_{t_{n}}^{t}du\left( h_{u}^{-1}L_{u}\left( h_{u}\right)
+h_{u}^{-1}\partial _{u}h_{u}\right) \right) \right] h_{t}(y) \\
&=&\mathbf{E}_{s,x}\left[ F(x_{s},x_{t_{1}},x_{t_{2}},...,x_{t_{n}},x_{t})h_{s}^{-1}(x_{s})\exp
\left( -\int_{s}^{t}du\left( h_{u}^{-1}L_{u}\left( h_{u}\right)
+h_{u}^{-1}\partial _{u}h_{u}\right) \right) h_{t}(x_{t})\right] \\
&=&\mathbf{E}_{s,x}\left[ F_{\left[ s,t\right] }\exp (-\mathbf{Z}_{s}^{h,t}\left[ x
\right] )\right].
\end{eqnarray*}
We thus arrive at (\ref{martex}).


\begin{thebibliography}{99}

\bibitem{Agarw}  Agarwal, G. S., \textit{Fluctuation-dissipation theorems
for systems in non-thermal equilibrium and applications}. Z. Physik \textbf{
252}, 25-38 (1972).

\bibitem{App}  Applebaum, D. , Levy Process and Stochastic Calculus.
Cambridge University Press, (2004).

\bibitem{Bae1}  Baiesi, M., Maes, C., Wynants, B. , \textit{Fluctuations and
response of nonequilibrium states}, Phys. Rev. Lett. \textbf{103}, 010602 (2009).

\bibitem{Bae2}  Baiesi, M., Maes, C., Wynants, B. , \textit{Nonequilibrium
linear response for Markov dynamics, I: jump processes and overdamped
diffusions}. J. Stat. Phys. \textbf{137}, 1094-1116 (2009).

\bibitem{Bas1} Bass, R., SDEs with Jumps, Notes for Cornell Summer School. http://www.math.uconn.edu/~bass/cornell.pdf, (2007).

\bibitem{Bau1}  Baule, A., Cohen, E. G. D. , \textit{Fluctuation properties
of an effective nonlinear system subject to Poisson noise}. Phys. Rev. E \textbf{79}, 030103 (2009).

\bibitem{Blu1}  Blumenthal, R.M., Getoor, R.K. , Markov processes and potential theory. Academic Press (1968).

\bibitem{Bri}  Brissaud, A., Frisch, U. , \textit{Linear Stochastic
differential equation}. J. Math. Phys. \textbf{15}, 5 (1974).

\bibitem{Call1}  Callen, H.B., Welton, T. A. , \textit{Irreversibility and
generalized noise}. Phys. Rev. \textbf{83}, 34-40 (1951).

\bibitem{CHCHJAR}  Chernyak, V., Chertkov, M., Jarzynski, C. , \textit{
Path-integral analysis of fluctuation theorems for general Langevin
processes }. J. Stat. Mech. P08001 (2006). 


\bibitem{Che1}  Chetrite, R., Gawedzki, K. , \textit{Fluctuation relations
for diffusion processes}. Commun. Math. Phys. \textbf{282}, 469-518 (2008).

\bibitem{Fal2}  Chetrite, R., Falkovich, G., Gawedzki, K. , \textit{
Fluctuation relations in simple examples of nonequilibrium steady states}.
J. Stat. Mech. P08005 (2008).

\bibitem{Che3}  Chetrite, R., Gawedzki, K. , \textit{Eulerian and Lagrangian
pictures of nonequilibrium diffusions}. J. Stat. Phys \textbf{137}, 5-6
(2009)

\bibitem{Che4}  Chetrite, R. , \textit{Fluctuation relations for diffusion
that is thermally driven by a nonstationary bath}. Phys. Rev E. 051107 (2009)

\bibitem{Che5}  Chetrite, R. , Thesis of ENS-Lyon (2008). Manuscript
available at \textit{http://perso.ens-lyon.fr/raphael.chetrite/}

\bibitem{Chu1}  Chung, K.L., Walsh, J.B. , Markov Processes, Brownian
Motion, and Time Symmetry. Springer Science, Second Edition (2005)

\bibitem{Con}  Cont, R., Tankov, P. , Financial modeling with Jump
Processes. Chapman Hall (2003)

\bibitem{Cri1}  Crisanti, A., Ritort, J. , \textit{Violation of the
fluctuation-dissipation theorem in glassy systems: basic notions and the
numerical evidence}. J. Phys. A: Math. Gen. \textbf{36}, R181-R290
 (2003).
\bibitem{Crooks2}  Crooks, G.E. , \textit{Path ensembles averages in systems
driven far from equilibrium}. Phys. Rev. E \textbf{61}, 2361-2366 (2000).

\bibitem{Cug1}  Cugliandolo, L.F., Kurchan, J., Parisi, G. , \textit{Off
equilibrium dynamics and aging in unfrustrated systems}. J. Physique \textbf{
4}, 1641-1656 (1994).

\bibitem{Cze1}  Czernik, T., Kula, J., Luczka, J., Hanggi, P. , \textit{
Thermal ratchets driven by Poissonian white shot noise.}Phys. Rev. E 
\textbf{55}, 4057-4066 (1997).

\bibitem{Dar1}  Darses, S. , Nourdin, I., \textit{Dynamical properties and
characterization of gradient drift diffusion}. Elec. Comm. Probab. \textbf{12
}, 390-400 (2007).

\bibitem{Dav1}  Davis, M.H.A., \textit{Piecewise-deterministic Markov
Process: A General Class of Non-Diffusion Stochastic Models}. J.R. Statist.
Soc. B \textbf{46}, 353-388 (1984).

\bibitem{Del}  Dellacherie, C., Meyer, P.A. : Probabilit\'{e}s et potential. Chapitre V \`{a} VIII , volume \textbf{1385} of Actualit\'{e}s Scientifiques et Industrielles. Hermann 1980.  

\bibitem{Dem1}  Dembo, A., Deuschel, J.D., \textit{Markovian perturbation,
response and fluctuation dissipation theorem}. To appear in Ann. Inst. Henri
Poinc. (2010).

\bibitem{Dia1} Diaconis, P., Miclo, L., \textit{On characterizations of
Metropolis type algorithms in continuous time}. Alea  \textbf{6},199-238 (2009).

\bibitem{Die1}  Diezemann, G., \textit{Fluctuation-dissipation relations
for Markov processes}. Phys. Rev. E \textbf{72}, 011104 (2005).

\bibitem{Doo1}  Doob, J.L., Classical Potential Theory and Its
Probabilistic Counterpart. Springer-Verlag, New York (1984).

\bibitem{Dur}  Durrett, R., Probability, Theory and Examples. Fourth Edition.  Cambridge University Press. 2010. 

\bibitem{Dyn1}  Dynkin, E.B., \textit{The initial and final behavior of
trajectories of Markov Processes}. Russ. Math. Surv. \textbf{26}, 165 (1971).

\bibitem{Dyn2}  Dynkin, E.B., \textit{On duality for Markov processes}, in
\textit{Stochastic Analysis}. Ed A. Friddman, M. Pinsky, Academic Press (1978).

\bibitem{Eva}  Evans, D.J., Searles, D.J. , \textit{Equilibrium microstates
which generate second law violating steady states}. Phys. Rev. E \textbf{50}, 1645-1648
(1994).

\bibitem{Fel1}  Feller, W. , \textit{On the integro-differential equations
of purely discontinuous markoff processes}. Trans Am Math Soc \textbf{48}, 488-515 (1940).

\bibitem{Fol1}  Follmer, H. , \textit{\ An entropy approach to the time
reversal of diffusion process.} In Stochastic differential equation 156-163,
Lecture Notes in Control and Information Sci., 69, Springer, 1985.

\bibitem{Gal}  Gallavotti, G., Cohen, E.G.D. , \textit{Dynamical ensemble in
a stationary state}. J. Stat. Phys. \textbf{80}, 931-970 (1995).

\bibitem{Gall2}Gallavotti, G., \textit{Extension of Onsager's reciprocity to large fields and the chaotic hypothesis.} Phys. Rev. Lett. \textbf{77}, 4334-4337 (1996).

\bibitem{Ge}Ge, H., Jiang, D.Q., \textit{Generalized Jarzynski's equality of inhomogeneous multidimensional diffusion processes}. J.Stat. Phys. \textbf{131}, 675-689 (2008).

\bibitem{Gom1}Gomez-Solano, J.R., Petrosyan, A., Ciliberto, S., Chetrite, R., Gawedzki, K., \textit{Experimental verification of a modified fluctuation-dissipation relation for a micron-sized particle in a nonequilibrium steady state}. Phys. Rev. Lett. \textbf{103}, 040601 (2009).

\bibitem{Gom2}Gomez-Solano, J.R., Petrosyan, A., Ciliberto, S. Maes, C., \textit{Non-equilibrium linear response of micron-sized systems.} arXiv:1006.3196v1.

\bibitem{Han2}Hanggi, P., \textit{Langevin description of Markovian integro-differential master equations.} Z Physik B \textbf{36}, 271-282 (1980). 

\bibitem{Han}Hanggi, P., Thomas, H. ,\textit{Stochastic processes: time evolution, symmetries and linear response}. Physics Reports \textbf{88}, 207-319 (1982).

\bibitem{Sch1}Harris, R.J., Sch\"{u}tz, G.M., \textit{Fluctuation theorems for stochastic dynamics}. J. Stat. Mech., P07020 (2007).

\bibitem{HatSas}  Hatano, T., Sasa, S. , \textit{Steady-state thermodynamics
of Langevin systems}. Phys. Rev. Lett. \textbf{86}, 3463-3466 (2001).

\bibitem{Hau1}  Haussmann, U.G., Pardoux, E. , \textit{Time reversal of
diffusions}. The Annals of Probability, \textbf{14}, 1188-1205 (1986).

\bibitem{Ito}  Ito, K., Watanabe, S. , \textit{Transformation of Markov
processes by multiplicative functionals}. Ann. Inst. Fourier. \textbf{15}, 13-30 (1965)

\bibitem{Jac1}  Jacob, N. , Pseudo-differential Operators and Markov
Processes. Vol I, II, III Imperial College Press, London, (2001), (2002), (2005).

\bibitem{Jarz}  Jarzynski, C., \textit{A nonequilibrium equality for free
energy differences}. Phys. Rev. Lett. \textbf{78}, 2690-2693 (1997).

\bibitem{Jarz2}  Jarzynski, C., \textit{Equilibrium free energy differences
from nonequilibrium measurements: a master equation approach}. Phys. Rev. E 
\textbf{56}, 5018 (1997).

\bibitem{Jia1}  Jiang, D.Q., Qian, M., Qian, M.P , Mathematical theory of
nonequilibrium steady states : on the frontier of probability and dynamical
systems. Lecture notes in mathematics \textbf{1833}, (2004).

\bibitem{Joa} Joachain, C. : Quantum collision theory, North-Holland Publishing, 1975.

\bibitem{Kubo0}  Kubo, R. , \textit{The fluctuation-dissipation theorem}.
Rep. Prog. Phys. \textbf{29}, 255-284 (1966).

\bibitem{Kubo2} Kubo, R., Toda, M., Hashitsume, N., Statistical Physics II: Non-equilibrium Statistical Mechanics, Springer, (1991). 

\bibitem{Kil1}  Klimontovich, Yu.K. , \textit{Ito, Statonovich and kinetic
forms of stochastic equations}. Physics A \textbf{163}, 515-532 (1990).

\bibitem{Kun1}  Kunita, H. , \textit{Absolute continuity of Markov process
and generators}. Nagoya Math. J. \textbf{36}, 1-26 (1969).

\bibitem{Kurchan}  Kurchan, J. , \textit{Fluctuation theorem for stochastic
dynamics}. J. Phys. A: Math. Gen. \textbf{31}, 3719-3729 (1998).

\bibitem{Kurchan2}  Kurchan, J. , \textit{Non-equilibrium work relations}.
J. Stat. Mech.: Theory Exp., P07005 (2007).

\bibitem{LebowSp}  Lebowitz, J., Spohn, H. , \textit{A Gallavotti-Cohen type
symmetry in the large deviation functional for stochastic dynamics}. J.
Stat. Phys. \textbf{95}, 333-365 (1999).

\bibitem{Lip1}  Lippiello, E. , Corberi, F. , Zannetti , \textit{
Off-equilibrium generalization of the fluctuation dissipation theorem for
Ising spins and measurement of the linear response function}. Phys Rev E 
\textbf{71}, 036104 (2005).

\bibitem{Lip2}  Lippiello, E. , Corberi, F., Sarracinno, A., Zannetti , \textit{
Non-linear response and fluctuation dissipation relations}. Phys Rev E 
\textbf{78}, 041120 (2008).

\bibitem{Liu1}  Liu, F., Ou-Yang, Z.C., \textit{Generalized integral
fluctuation theorem for diffusion processes}. Phys. Rev. E. \textbf{79},
060107 (2009)

\bibitem{Liu3}  Liu, F. , Luo, Y.P., Huang, M.C., Ou-Yang, Z.C. , \textit{A
generalized integral fluctuation theorem for general jump processes}. J.
Phys. A: Math. Theor. \textbf{42}, 332003 (2009).

\bibitem{Liu2}  Liu, F., Ou-Yang, Z.C. , \textit{Linear response theory and
transient fluctuation-theorems for diffusion processes: a backward point of
view}. arXiv:0912.1917v1.

\bibitem{Luc1}  Luczka, J., Czernik, T., Hanggi, P. , \textit{Symmetric
white noise can induce directed current in ratchets.} Phys. Rev. E \textbf{56} 3968-3975 (1997).

\bibitem{Maes}  Maes, C. , \textit{The Fluctuation Theorem as a Gibbs Property
}. J. Stat. Phys. \textbf{95}, 367-392 (1999).

\bibitem{Mae3}  Maes, C., Wynants, B. , \textit{On a response formula and
its interpretation}. Markov Processes and Related Fields \textbf{16}, 45-58 (2010).

\bibitem{MPRV}  Marini Bettolo Marconi, U., Puglisi, A., Rondoni. L.,
Vulpiani, A. , \textit{Fluctuation-dissipation: response theory in
statistical physics}. Phys. Rep. \textbf{461}, 111-195 (2008).

\bibitem{May1}  Mayer, P., Leonard, S., Berthier, L., Garrahan, J.P.,
Sollich, P. , \textit{Activated Aging Dynamics and Negative
Fluctuation-Dissipation Ratios}. Phys. Rev. Lett. \textbf{96}, 030602 (2006).

\bibitem{Mer1}  Merton, R. , \textit{Option pricing when underlying stock
returns are discontinuous}. Journal of Financial Economics \textbf{3}, 125-144
(1976).


\bibitem{Nel}  Nelson, E. , Dynamical Theories of Brownian Motion, second
edition (2001), Princeton University Press. (1967).

\bibitem{Nel2}  Nelson, E. , Quantum Fluctuations, Princeton Series in
Physics, Princeton University Press, (1985).

\bibitem{Ons1}  Onsager, L. , \textit{Reciprocal relations in irreversible
processes I}. Phys. Rev. \textbf{37}, 405-426 (1931).

\bibitem{Ons2}  Onsager, L. , \textit{Reciprocal relations in irreversible
processes II}. Phys. Rev. \textbf{38}, 2265-2279 (1931).

\bibitem{Oon1}  Oono, Y., Paniconi, M. , \textit{Steady State Thermodynamics}.
 Prog. Theor. Phys. Suppl. \textbf{130}, 29-44 (1998).

\bibitem{Pal1}  Palmowski, Z., Rolski, T. , \textit{A Technique for
exponential change of measure for Markov processes}. Bernoulli \textbf{8}, 6 (2002).
767-785.

\bibitem{Por1}  Porporato, A., D'Odorico, P. , \textit{Phase Transitions
Driven by State-Dependent Poisson Noise. }Phys. Rev. Lett., \textbf{92}, 110601
(2004).

\bibitem{Pro1}  Prost, J., Joanny, J.F. , Parrondo, J.M.R. , \textit{
Generalized fluctuation-dissipation theorem for steady state systems}. Phys
Rev Lett \textbf{103}, 090601 (2009).

\bibitem{Rev1}  Revuz, D., Yor, M. , Continuous martingales and Brownian
Motion. Springer-Verlag, Berlin Third edition (1999).

\bibitem{Risk}  Risken, H. , The Fokker Planck Equation. 2$^{\mathrm{nd}}$
Edition. Springer, Berlin-Heidelberg (1989)

\bibitem{San1}  Sancho, J.M., San Miguel, M., Pesquera, L. , Rodriguez, M.
A. , \textit{Positivity requirements on fluctuating parameters}. Physica A 
\textbf{142}, 532-547 (1987).

\bibitem{Sei}  Seifert, U. , \textit{Stochastic thermodynamics: Principles
and perspectives}. Eur. Phys. Journ. B \textbf{64}, 423-432 (2008).

\bibitem{Sei2}  Seifert, U., Speck, T. , \textit{\ Fluctuation-dissipation
theorem in nonequilibrium steady states}. Europhys.Lett. \textbf{89}, 10007 (2010).


\bibitem{Sek}  Sekimoto, K. ,  Stochastic Energetics. Lecture Notes in Physics \textbf{799}, Springer Verlag. 2010.


\bibitem{SpS0}  Speck, T., Seifert, U , \textit{Restoring a
fluctuation-dissipation theorem in a nonequilibrium steady state}. Europhys.
Lett. \textbf{74}, 391-396 (2006).

\bibitem{Spe2}  Speck, T. , Driven Soft Matter: \textit{Entropy Production
and the Fluctuation-Dissipation Theorem}. arXiv:1004.1621.

\bibitem{Str}  Stroock, D., Varadhan, S. R. S. , Multidimensional Diffusion
Processes. Springer, Berlin (1979)

\bibitem{Str2}  Stroock, D. , \textit{Diffusion processes associated with
L\'{e}vy generators. }Probability Theory and Related Fields \textbf{32}, 209-244
(1975).

\bibitem{Tou1}  Touchette, H., Cohen, E. G. D. ,\ \textit{Fluctuation
relation for a Levy particle}. Phys. Rev. E \textbf{76}, 020101 (2007)

\bibitem{Van2}  Van Kampen, N.G. , \textit{Processes with delta-correlated
cumulants}. Physica A \textbf{102}, 489-495 (1980).



\bibitem{Zim1}  Zimmer, M.F. , \textit{Fluctuations in Nonequilibrium
Systems and Broken Supersymmetry}. J. Stat. Phys \textbf{73}, 751-764 (1993).
\end{thebibliography}
\end{document}